\documentclass[%
superscriptaddress,
reprint,
amssymb, 
amsfonts,
amsmath,
aps, 
prx,%
floatfix,
]{revtex4-2}
\usepackage{graphicx}%
\usepackage{dcolumn}%
\usepackage{bm}%

\usepackage[utf8]{inputenc}
\usepackage[T1]{fontenc}

\usepackage{braket}
\usepackage{upgreek}
\DeclareMathOperator{\Tr}{tr}
\DeclareMathOperator{\E}{E}

\usepackage{tikz}
\begin{document}

\title{Enhancing Coherence with a Clock Transition and Dynamical Decoupling in the Cr$_7$Mn Molecular Nanomagnet}
\author{Guanchu Chen}%
\affiliation{Department of Physics and Astronomy, Amherst College, Amherst, MA 01002, USA}
\affiliation{Department of Physics, University of Massachusetts Amherst, Amherst, MA 01003, USA}%

\author{Brendan C.~Sheehan}%
\altaffiliation{Current address: Department of Physics and Astronomy, Dartmouth College, Hanover, NH 03755, USA}
\affiliation{Department of Physics and Astronomy, Amherst College, Amherst, MA 01002, USA}
\affiliation{Department of Physics, University of Massachusetts Amherst, Amherst, MA 01003, USA}%

\author{Ilija Nikolov}%
\altaffiliation{Current address: Department of Physics, Brown University, Providence, RI 02912, USA}
\affiliation{Department of Physics and Astronomy, Amherst College, Amherst, MA 01002, USA}

\author{James W.~Logan}%
\altaffiliation{Current address: Department of Physics and Astronomy, Dartmouth College, Hanover, NH 03755, USA}
\affiliation{Department of Physics and Astronomy, Amherst College, Amherst, MA 01002, USA}

\author{Charles A.~Collett}%
\altaffiliation{Current address: Department of Physics, Hamilton College, Clinton, NY 13323, USA}
\affiliation{Department of Physics and Astronomy, Amherst College, Amherst, MA 01002, USA}

\author{Gajadhar Joshi}%
\altaffiliation{Current address: Center for Integrated Nanotechnologies, Sandia National Laboratories, Albuquerque, NM 87123, USA}
\affiliation{Department of Physics and Astronomy, Amherst College, Amherst, MA 01002, USA}

\author{Grigore A.~Timco}
\affiliation{Department of Chemistry, The University of Manchester, Manchester M13 9PL, UK}

\author{Jillian E.~Denhardt}
\altaffiliation{Current address: University of Hawai`i at Mānoa, Honolulu, HI 96822, USA}
\affiliation{Department of Chemistry, University of Massachusetts, Amherst, MA 01003, USA}

\author{Kevin R.~Kittilstved}
\altaffiliation{Current address: Department of Chemistry, Washington State University, Pullman, 99164-4630, USA}
\affiliation{Department of Chemistry, University of Massachusetts, Amherst, MA 01003, USA}

\author{Richard E.~P.~Winpenny}
\affiliation{Department of Chemistry, The University of Manchester, Manchester M13 9PL, UK}

\author{Jonathan R.~Friedman}%
\email{jrfriedman@amherst.edu}
\affiliation{Department of Physics and Astronomy, Amherst College, Amherst, MA 01002, USA}
\affiliation{Department of Physics, University of Massachusetts Amherst, Amherst, MA 01003, USA}

\date{July 18, 2025}

\begin{abstract}
Molecular magnets are attractive as spin qubits due to their chemical tunability, addressability through electron-spin resonance techniques, and long coherence times. Clock transitions (CTs), for which the system is immune to the effect of magnetic-field fluctuations to first order, provide a method to enhance the coherence time $T_2$, and to reveal mechanisms of decoherence that are not due to such fluctuations. Here we investigate two variants of Cr$_7$Mn, a spin-1 molecular nanomagnet, at fields near a zero-field CT. We find that at temperatures $\le2$~K, $T_2\sim1\,\upmu$s at the CT using a Hahn-echo pulse sequence.  Away from the CT, electron-spin-echo envelope modulation (ESEEM) oscillations due to coupling to nuclear spins are observed and have a $T_2$ as high as $1.35$\,$\upmu$s, indicating a distinct mechanism of coherence preservation.  Dynamical decoupling with the CPMG pulse sequence yields $T_2\sim\!2.8~\upmu$s at the CT and up to $\sim\!3.6\,\upmu$s in the ESEEM regime along with a demodulation of the oscillatory behavior. The experimental values of $T_2$ are largely independent of the degree of dilution of the molecules in solvent or whether the solvent is deuterated, indicating that much of the decoherence and ESEEM arises from sources within the molecules themselves. To account for decoherence, we develop a model that includes not only field fluctuations but also fluctuations in the CT transition frequency itself. Our results can be well explained by treating the environment as a combination of noise at the nuclear Larmor precession frequency and $1/f$ noise in the transverse anisotropy parameter $E$. Such information about the microscopic origins of decoherence can aid the rational design of molecular-based spin qubits. 
\end{abstract}
\maketitle

\section{Introduction}
At the core of quantum information science is the ability to manipulate a qubit -- a two-level system that can exist in superposition of states for a nontrivial duration of time. Many physical systems have exhibited characteristics that make  them attractive as qubits, including superconducting circuits~\cite{kjaergaardSuperconductingQubitsCurrent2020}, trapped ions~\cite{bruzewiczTrappedionQuantumComputing2019}, and spin qubits~\cite{mortonEmbracingQuantumLimit2011,heinrichQuantumcoherentNanoscience2021,coronadoMolecularMagnetismChemical2019,chiesaBlueprintMolecularSpinQuantum2023,gaita-arinoMolecularSpinsQuantum2019,wesenbergQuantumComputingElectron2009}. Molecular nanomagnets (MNMs), paramagnetic molecules with a ground-state magnetic moment, are particularly interesting as spin qubits due to their chemically engineerable electronic spin states, which can be probed through the use of electron spin resonance (ESR) techniques~\cite{friedmanSingleMoleculeNanomagnets2010,leuenbergerQuantumComputingMolecular2001,tejadaMagneticQubitsHardware2001,luisObservationQuantumCoherence2000,chiesaMolecularNanomagnetsViable2024}. To be considered a viable qubit, an MNM must retain phase memory for times long compared to quantum gate operation times. 

One way to achieve a long phase-memory (coherence) time $T_2$ is through a so-called clock transition (CT), which occurs at an avoided level crossing where the transition frequency $\epsilon$ between states is immune to the decohering effects of environmental noise to first order. CTs are commonly used in superconducting qubits, particularly the transmon and fluxonium, to provide isolation from fluctuations from charge and/or magnetic noise sources~\cite{vionManipulatingQuantumState2002,devoretSuperconductingCircuitsQuantum2013,krantzQuantumEngineersGuide2019,manucharyanFluxoniumSingleCooperPair2009,somoroffMillisecondCoherenceSuperconducting2023}. In spin systems, where decoherence is primarily due to magnetic-field fluctuations from environmental spins, a CT occurs when $\nabla_{\boldsymbol{B}} \epsilon = \mathbf{0}$; when this condition is fulfilled, $T_2$ is typically significantly enhanced. CTs  have been observed in several spins systems, such as in Bi-doped silicon~\cite{wolfowiczAtomicClockTransitions2013}, the HoW$_{10}$ molecular magnet~\cite{shiddiqEnhancingCoherenceMolecular2016} and other MNMs~\cite{collettClockTransitionCr7Mn2019,kundu92GHzClockTransition2022,zadroznyPorousArrayClock2017,rubin-osanzChemicalTuningSpin2021,gakiya-teruya546GHzClock2025,tlemsaniAssessingRobustnessClock2025,gimenez-santamarinaExploitingClockTransitions2020,lewisClockTransitionsGuard2021}, as well as spin defects in silica glasses~\cite{sheehanClockTransitionsGenerated2024}. In addition, theoretical work has proposed implementations of one- and two-qubit gates in an MNM dimer in which all of the relevant radiative transitions are CTs~\cite{collettConstructingClocktransitionbasedTwoqubit2020}. MNMs provide excellent testbeds for studying spin CTs because their properties can be chemical engineered. The local molecular environment can be adjusted via ligand-field tuning, isotopic purification, or choice of matrix (e.g.~solvent), allowing control of the dominant decoherence channel for the system~\cite{laorenzaTunableCr4Molecular2021,baylissEnhancingSpinCoherence2022,hendersonControlInhomogeneityDegree2008,wedgeChemicalEngineeringMolecular2012}. 

While CTs typically enhance $T_2$, portending better qubit performance, they also provide a window into the underlying decoherence mechanisms in a system. Since at a CT decoherence from magnetic-field fluctuations is suppressed, the remaining decoherence may arise from other sources, e.g.~vibrations~\cite{mondalSpinphononDecoherenceSolidstate2023,luisSpinlatticeRelaxationQuantum2010,ullahSilicoMolecularEngineering2019}, that are not filtered by the CT. When one tunes away from the CT by changing the applied field, the system is more susceptible to decoherence from fluctuating fields and the change in coherence reflects the nature of these fluctuations. Gaining understanding about the sources of decoherence in spin qubits can thus provide fundamental information and aid in the rational design of the next generation of qubits. 

In this work, we employ pulse ESR to study decoherence in the  Cr$_7$Mn MNM near its zero-field CT~\cite{collettClockTransitionCr7Mn2019}. We find $T_2$ values on the order of a few $\upmu$s at the CT as well as in a range of fields where electron-spin-echo envelope modulation (ESEEM) oscillations are observed.  Dynamical decoupling techniques provide further enhancement to coherence in both of these regimes.  Through a combination of experimental studies  and detailed modeling of the dependence of decoherence on field and pulse-sequence properties, we extract information about the environmental sources of noise that give rise to decoherence in this system.  We find that the noise consists of field fluctuations arising from a nuclear spin bath combined with noise in the CT transition frequency itself.  While the CT can effectively filter out magnetic field noise, it cannot filter the latter form of noise.  This suggests that fluctuations in the CT frequency, arising from fluctuations in non-magnetic Hamiltonian parameters, may ultimately limit the efficacy of CTs in enhancing coherence.

\begin{figure}[!hb]
    \centering
    \includegraphics[width=\linewidth]{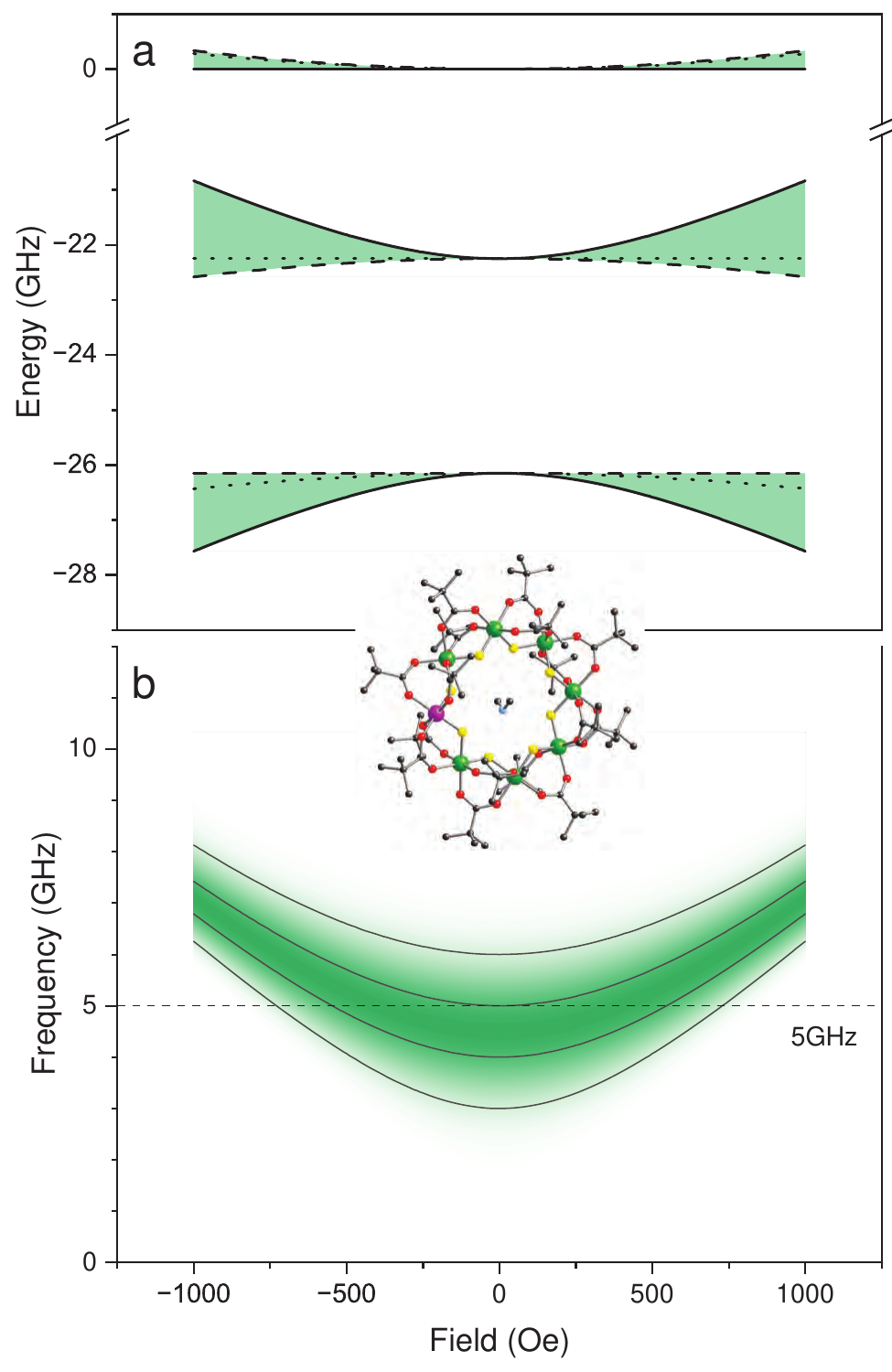}
    \caption{Avoided crossing at zero field with broadening. a) Eigenenergies of the two lowest levels of Cr$_7$Mn, calculated from Eq.~\ref{eqn:Cr7Mn_Ham}, as a function of field; solid, dashed, and dotted lines shows when the field is applied along the $z$, $x$, and $y$ axes, respectively. The colored areas illustrate the range of energies given all possible orientations of field. b) Transition frequency as a function of magnetic field calculated from Eq.~\ref{eqn:Cr7Mn_Ham} and assuming that the field is directed along the easy ($z$) axis. To illustrate the effect of inhomogeneous broadening, we let the transverse anisotropy parameter $E$ have a Gaussian distribution centered at $E=2.25$~GHz with a full width at half maximum of $1$~GHz. Color indicates probability density associated with this distribution at each field. The dashed line indicates an example probe frequency of $5$~GHz, illustrating that, because of the significant broadening in this system, resonance with a single radiation frequency occurs over a wide range of magnetic fields. Inset shows a ball-and-stick diagram of \textbf{1}. Colors: Cr$^{3+}$ (green), Mn$^{2+}$ (purple), F (yellow), O (red), C (black), N (light blue).  Hydrogens have been omitted for clarity. %
    } 
    \label{fig:levels}
\end{figure}

Cr$_7$Mn is one of a group of heterometallic rings described by A-[Cr$_7$MnF$_8$X$_{16}$], where A is the cation and X indicates the ligand -- in our samples, X = (CH$_3$)$_3$CCOO$^-$. In this work, we study two variants of Cr$_7$Mn that differ only in terms of cation: A = (CH$_3$)$_2$NH$_2^+$ (\textbf{1}) and A = Cs$^+$ (\textbf{2})~\cite{larsenSynthesisCharacterizationHeterometallic2003,wedgeChemicalEngineeringMolecular2012}. The structure of \textbf{1} is shown in the inset of Fig.~\ref{fig:levels}. These systems, and the structurally similar spin-$1/2$ Cr$_7$Ni, were previously reported to have $T_2$ ranging from a few hundred~ns up to $15~\upmu$s at temperatures below $5$~K in X-band~\cite{piligkosEPRSpectroscopyFamily2009,ardavanWillSpinRelaxationTimes2007,wedgeChemicalEngineeringMolecular2012}. While there has been much research on Cr$_7$Ni, ~\cite{ardavanWillSpinRelaxationTimes2007,timcoEngineeringCouplingMolecular2009,wedgeChemicalEngineeringMolecular2012,ardavanEngineeringCoherentInteractions2015,liuElectricFieldControl2019,chiesaMolecularNanomagnetsSwitchable2014,ferrando-soriaModularDesignMolecular2016}, which has $S=\frac{1}{2}$, Kramers' theorem prohibits it from exhibiting a zero-field CT. In contrast, Cr$_7$Mn has a ground-state spin of $S=1$~\cite{larsenSynthesisCharacterizationHeterometallic2003,timcoEngineeringCouplingMolecular2009,lancasterRelaxationMuonSpins2010}, the integer spin giving rise to a zero-field CT. 

Typically spin CTs arise from one of two mechanisms: 1) Strong hyperfine coupling between an electronic and nuclear spin~\cite{kundu92GHzClockTransition2022,kunduElectronnuclearDecouplingSpin2023,ngoLargeHyperfineCoupling2025} or 2) Magnetic anisotropy with a significant transverse component. The latter mechanism is at play in Cr$_7$Mn, which at low temperature can be described by the spin Hamiltonian
\begin{equation}
\label{eqn:Cr7Mn_Ham}
    \mathcal{H}=-D S_z^2+E (S_x^2-S_y^2)+g_s \mu_B\boldsymbol{B}\cdot\boldsymbol{S},
\end{equation}
where $D$ and $E$ are longitudinal and transverse anisotropy parameters, respectively. Previous work found the values of the anisotropy parameters to be $D=21$~GHz, $E=1.9$~GHz, and the Land\'e factor to be $g=1.96$ ~\cite{ardavanEngineeringCoherentInteractions2015,collettClockTransitionCr7Mn2019}, although there is substantial inhomogeneity in $E$, as discussed in more detail below. Working in the $\{\ket{-1},\ket{0},\ket{1}\}$ basis, the eigenbasis of $S_z$, the zero-field eigenstates of Eq.~{\ref{eqn:Cr7Mn_Ham}} are $\ket{\pm} = \frac{1}{\sqrt{2}}\left(\ket{1}\pm\ket{-1}\right)$ and $\ket{0}$ with energies $-D\pm E$ and $0$, respectively. Because $D \gg E$ and given the probe frequencies used in our experiments, the $\ket{0}$ state can be ignored, resulting in a effective two-level system consisting of the states $\ket{\pm}$. At zero magnetic field, an avoided crossing with tunnel splitting energy $\epsilon = 2E$ emerges; when a magnetic field is applied to the two-level qubit, the energy levels exhibit a weak field dependence. Fig.~\ref{fig:levels}(a) shows the level energies calculated as a function of field with the field oriented along the cardinal directions. As can be seen, an avoided crossing is present at zero field in this system regardless of the direction of the field. In other words, this system has the structure of a CT in which the transition frequency is independent of magnetic field to first order: $\left.\nabla_{\boldsymbol{B}}\epsilon\,\right\vert_{B=0} = \boldsymbol{0}$. The system therefore experiences substantial protection from the decohering effects of magnetic-field fluctuations, leading to an enhancement of quantum coherence and the magnitude of the dephasing time $T_2$.

At zero field, the transition between the $\ket{\pm}$ states is determined only by the anisotropy parameter $E$. However, we find a significant range of frequencies at which a zero-field transition is observed (see \cite{supp}), indicating a substantial inhomogeneity in the value of $E$. The effect of this inhomogeneity is schematically illustrated in Fig.~\ref{fig:levels}(b), where we assume the value of $E$ belongs to a Gaussian distribution. The figure shows the simulated transition frequencies as a function of the field oriented along the $z$ direction, with color indicating the population of molecules in the ensemble with a specific transition frequency at a given applied field. Each solid gray line represents a different example of the transition frequency as a function of magnetic field for a single $E$ value; the gray line with the lowest minimum at zero field has the smallest $E$. %
The dashed horizontal line illustrates how a single ESR frequency (of $5$~GHz) samples a wide range of transition fields and therefore a wide range of $E$ values. All of our experiments employ orientationally disordered samples, resulting in the probing of molecules with field components along the $x$ and $y$ directions. Since the transition frequency dependence is weaker for these directions (see Fig.~\ref{fig:levels}(a)) than for $z$, the resulting broadening of the resonance in field is even wider than illustrated in Fig.~\ref{fig:levels}(b).

\section{Methods}
Samples were synthesized using procedures published previously~\cite{piligkosEPRSpectroscopyFamily2009}.  In order to reduce the effects on intermolecular dipole interactions, we studied dilute samples using either solid or liquid solutions. %
For solid-solution samples, we co-crystalized Cr$_7$Mn with a diamagnetic isostructural analog, Ga$_7$Zn~\cite{moroCoherentElectronSpin2014,collettClockTransitionCr7Mn2019}. Liquid samples were made by dissolving the molecule in toluene and sealing the solution in a fused-silica capillary with a torch. For some samples, deuterated toluene was used as a solvent to mitigate potential decoherence from interactions with solvent protons. Dilution percentages indicated herein indicate volume fraction of Cr$_7$Mn molecules in solvent/matrix. We employed two different procedures to cool down the toluene-based samples: they were either gradually cooled down to the base temperature ($\sim\!1.8$~K) in a Quantum Design Physical Property Measurement System (PPMS) cryostat, or flash frozen in liquid nitrogen prior to being quickly inserted into a pre-cooled cryostat to prevent any thawing.

Despite using different cations, different host matrices, and different cooling techniques, all of our samples behave similarly across a broad frequency range (see \cite{supp} for comparisons). This suggests that the mechanisms of decoherence underlying our observations are rather distinct from what was investigated previously~\cite{ardavanWillSpinRelaxationTimes2007,wedgeChemicalEngineeringMolecular2012}, where ESR was performed at X-band at fields far from any CT.

ESR experiments were performed with the sample placed in a copper loop-gap resonator (LGR)~\cite{fronciszLoopgapResonatorNew1982} in a custom-designed probe with \textit{in situ} frequency- and coupling-tuning capability~\cite{joshiAdjustableCouplingSitu2020}. To maximize signal from the CT, all data was collected using parallel mode ESR: the LGR's $B_1$ field was parallel to the DC field $B_0$. CW reflection spectroscopy was performed via a direct-detection method (i.e.~no field modulation) by monitoring the LGR's resonance with a Keysight E5063A vector network analyzer; as the field was swept, a change in the quality factor of the total resonance provided a measure of sample response. For pulsed spectroscopy we employed a homodyne detection scheme wherein a Tabor Electronics SE5082 arbitrary waveform generator (AWG), placed in a circuit with commercially available microwave electronics components, was used to generate pulses. Spin echoes were amplified, downconverted with a mixer, and then recorded with a digital oscilloscope. Phase cycling was used for background subtraction. A typical LGR used in this study has $Q\sim2000$ at $\sim$ 1.8~K; for pulsed measurements we placed Eccosorb-brand microwave absorber inside the copper shield near the resonator to lower the $Q$ and preserve pulse-shape fidelity.

\section{Results}
\begin{figure}
    \centering
    \includegraphics[width=\linewidth]{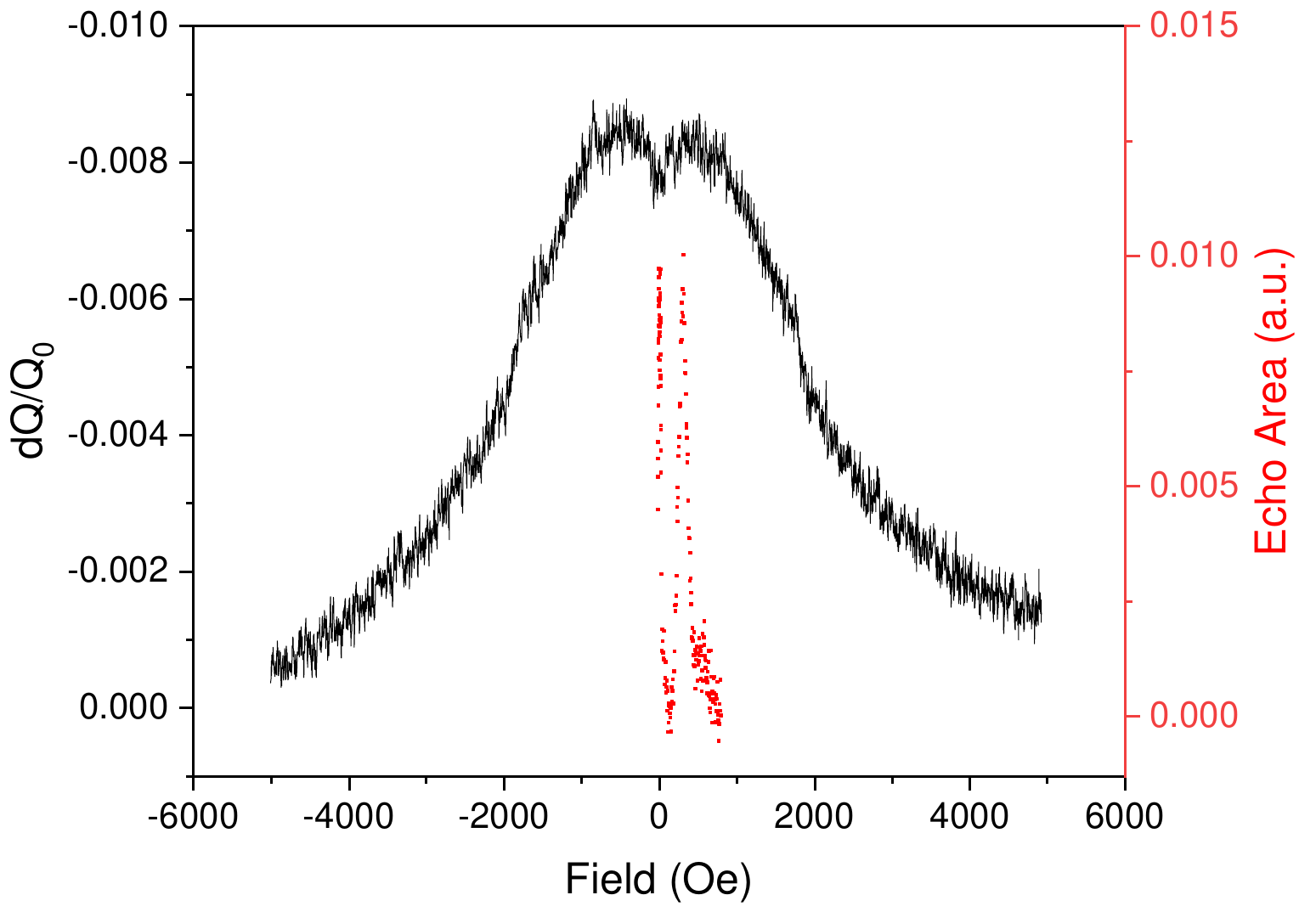}
    \caption{Spectra obtained from Cr$_7$Mn. Black shows the continuous-wave spectrum taken on a sample of 
    $5$\% dilution solid solution of \textbf{1} at $1.8$~K, using a radiation frequency of $5.013$~GHz. The fact that ESR signal is obtained over a rather wide range of field indicates the substantial inhomogeneous broadening in this system. Red shows the results of an echo-detected field spectrum on the same sample performed at $5.011$~GHz and at $1.8$~K using a pulse delay of $\tau=800$ ns.}
    \label{fig:cw}
\end{figure}

A typical CW spectrum of Cr$_7$Mn is shown in black in Fig.~\ref{fig:cw}, taken of a $5$\% dilution solid solution of \textbf{1}. The spectrum shows a broad spectral peak with a width of $\sim\!4000$~Oe. The large width suggests an inhomogeneous broadening of $\sim\!1$~GHz in the transverse anisotropy parameter $E$, indicating a ``softness'' to the structure of the molecule, perhaps due to a distortion of the ring. A hole-burning experiment indicates that despite the significant broadening in our samples, for given frequency and field, we are addressing a small sub-ensemble of the sample (details in \cite{supp}).

Also shown in Fig.~\ref{fig:cw} is an electron spin echo (ESE) spectra (red), which has signal over a much narrower range of fields than the CW spectrum: despite a CW signal at fields up to $\sim\!4000$~Oe, the echo signal is limited to fields below $1000$~Oe. This suggests that for most fields away from zero field, the coherence time $T_2$ is too short to allow for an echo to be observed.

\begin{figure}
    \centering
    \includegraphics[width=0.95\linewidth]{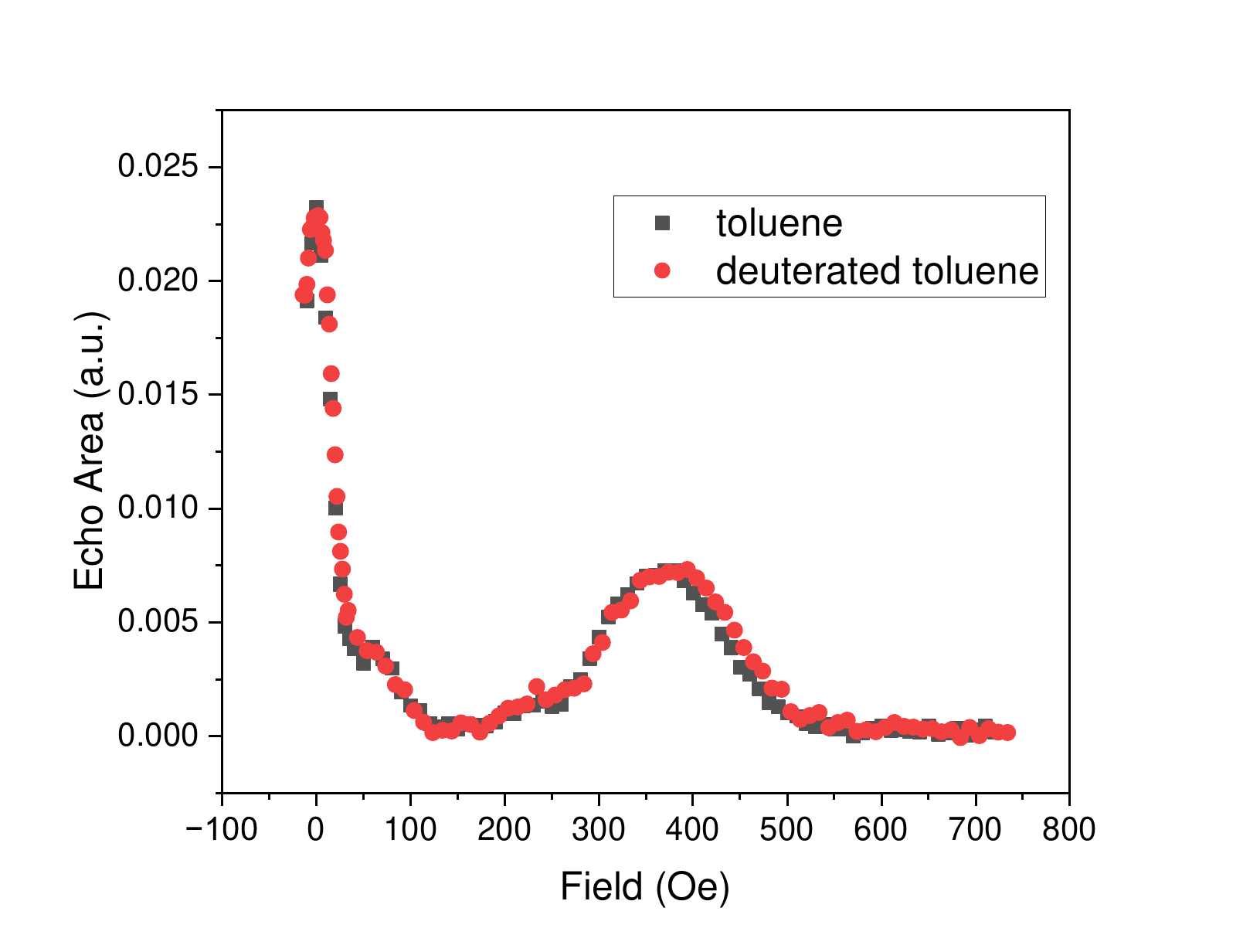}
    \caption{Echo-detected field dependence. The spectra were obtained from liquid solution samples of $1$\% dilution of \textbf{2} in toluene (black) and deuterated toluene (red)  at $1.8$~K, and $3916$~MHz and $3961$~MHz, respectively. The peaks at zero field and at $\sim390$~Oe are due to the CT and ESEEM, respectively. Both peaks are observed in spectra taken a various frequencies and are independent of the molecule's cation, the solvent used, or the cooling procedure. The delay time $\tau$ in the Hahn sequence used was $600$~ns. The position and height of the ESEEM peak changes with the value of $\tau$ employed, as discussed in the main text. 
    }
    \label{fig:EDFS}
\end{figure}

Fig.~\ref{fig:EDFS} shows echo-detected field-swept (EDFS) spectra from $1$\% dilution of \textbf{2} in toluene (black) or deuterated toluene (red) measured at $1.8$~K at similar frequencies. The spectra were obtained with a Hahn sequence with a delay time of $\tau=600$~ns. Two ESE peaks are visible; a narrow peak at zero field and a broad peak at higher field (in this case, $\sim390$~Oe). The narrow peak at zero field we attribute to a CT~\cite{collettClockTransitionCr7Mn2019}. The broader peak is the effect of ESEEM due to interactions with nuclear spins, as will be discussed in more detail below. Notably, we see very little difference between these two spectra, suggesting a limited role played by the solvent protons in producing the ESEEM behavior. We note the change in the amplitude of the ESEEM peak relative to the CT peak in Fig.~\ref{fig:EDFS} compared to Fig.~\ref{fig:cw}, which is due to the different values of $\tau$ used to acquire the data in each figure.

\begin{figure}
    \centering
    \includegraphics[width=\linewidth]{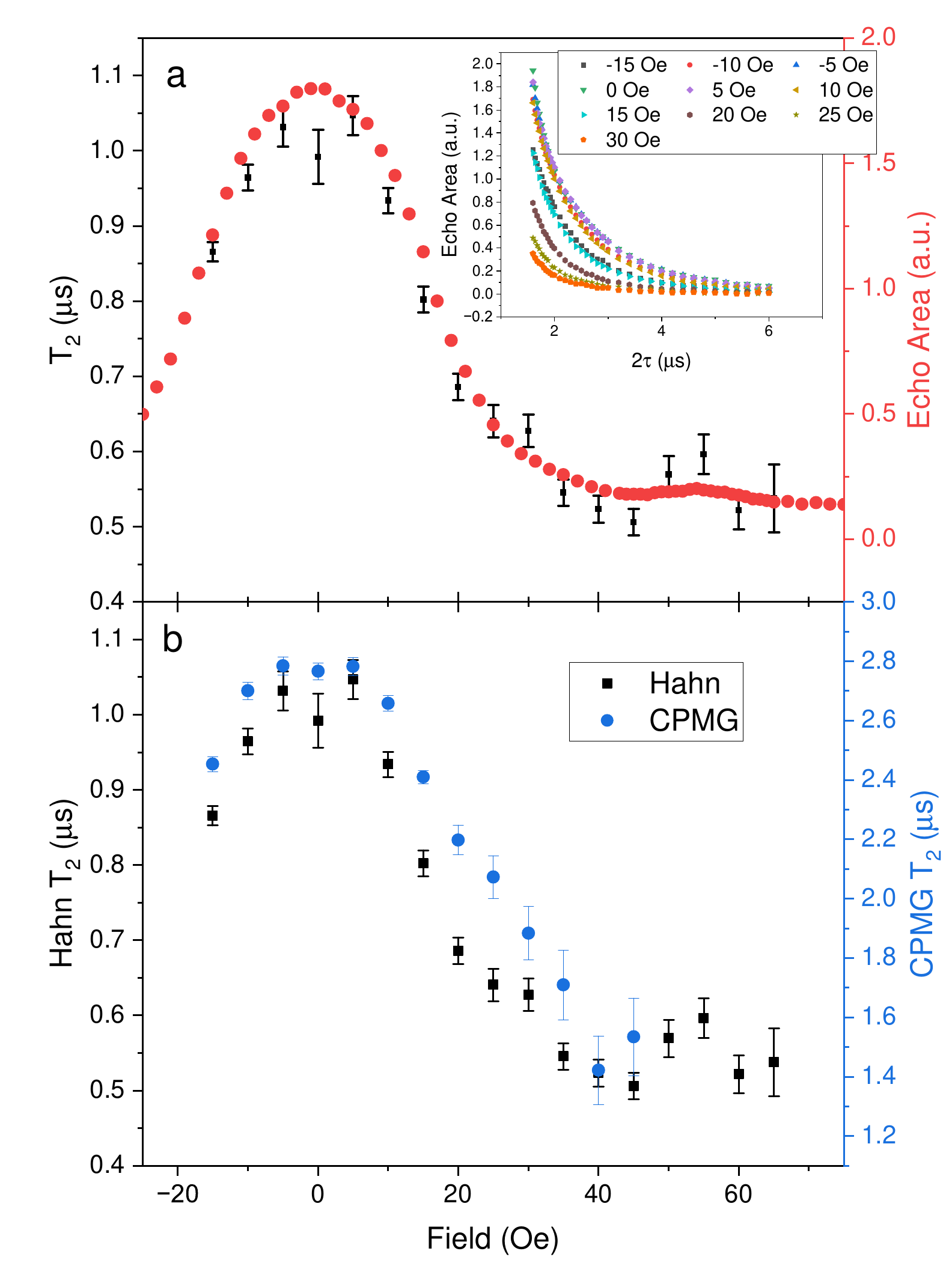}
    \caption{$T_2$ values near the CT. a) EDFS data (red) and Hahn $T_2$ (black). The inset shows echo area as a function of $2\tau$ for several values of magnetic field, as indicated. Fitting these curves to an exponential decay yields the $T_2$ values given in the main portion of the panel. The similarity in shape of the field dependence of the signal size and $T_2$ can be attributed to the reduction in $T_2$ as the field tunes the system away from the CT. The maximum value of $T_2$ is found to be $1.05\left(3\right)\,\upmu$s. b) $T_2$ measured with the CPMG pulse sequence (blue) overlaid with the values found using the Hahn sequence (black). The CPMG method yields values of $T_2$ $\sim\!2.8$ times larger than those found using the Hahn sequence. The delay time $\tau$ used in the CPMG sequence was $800$~ns. Data was taken with $5$\% dilution of \textbf{2} in toluene at $1.9$~K and $4639$~MHz.}
    \label{fig:CT_T2}
\end{figure}

Focusing on the CT peak, Fig.~\ref{fig:CT_T2}(a) shows the EDFS spectrum (red) overlaid with the coherence time $T_2$ (black) extracted from Hahn-echo data with a $5$\% dilution sample of \textbf{2}. Each $T_2$ value was extracted by varying the delay time $\tau$ in the Hahn sequence (Fig.~\ref{fig:CT_T2}(a) inset) and fitting the resulting decay to an exponential decay. The comparison in Fig.~\ref{fig:CT_T2}(a) shows that echo signal size and $T_2$ have similar behavior with field, indicating that the loss of echo signal with increasing field can be attributed to the reduction in $T_2$ as the field is applied and the system is tuned away from the CT. At the CT, we find a maximum coherence time of $1.05\left(3\right)~\upmu$s.

$T_2$ can be enhanced by employing the Carr-Purcell-Meiboom-Gill (CPMG) pulse sequence, whereby the spins are repeatedly refocused via repeated $\uppi$ pulses with uniform spacing $2\tau$, producing echoes at time $\tau$ after each $\uppi$ pulse. Each progressive echo area is smaller than the previous echo area; the exponential decay of the echoes during the CPMG sequence allows us to extract a value of $T_2$. Fig.~\ref{fig:CT_T2}(b) shows an overlay of $T_2$ measured with a Hahn sequence (black) and a CPMG sequence (blue), with separate axis scales for each set of data. Both methods show a similar behavior in how the coherence time changes with field around the CT, but the CPMG method produces $T_2$ values that are approximately 2.8 times larger.

\begin{figure}
    \centering
    \includegraphics[width=\linewidth]{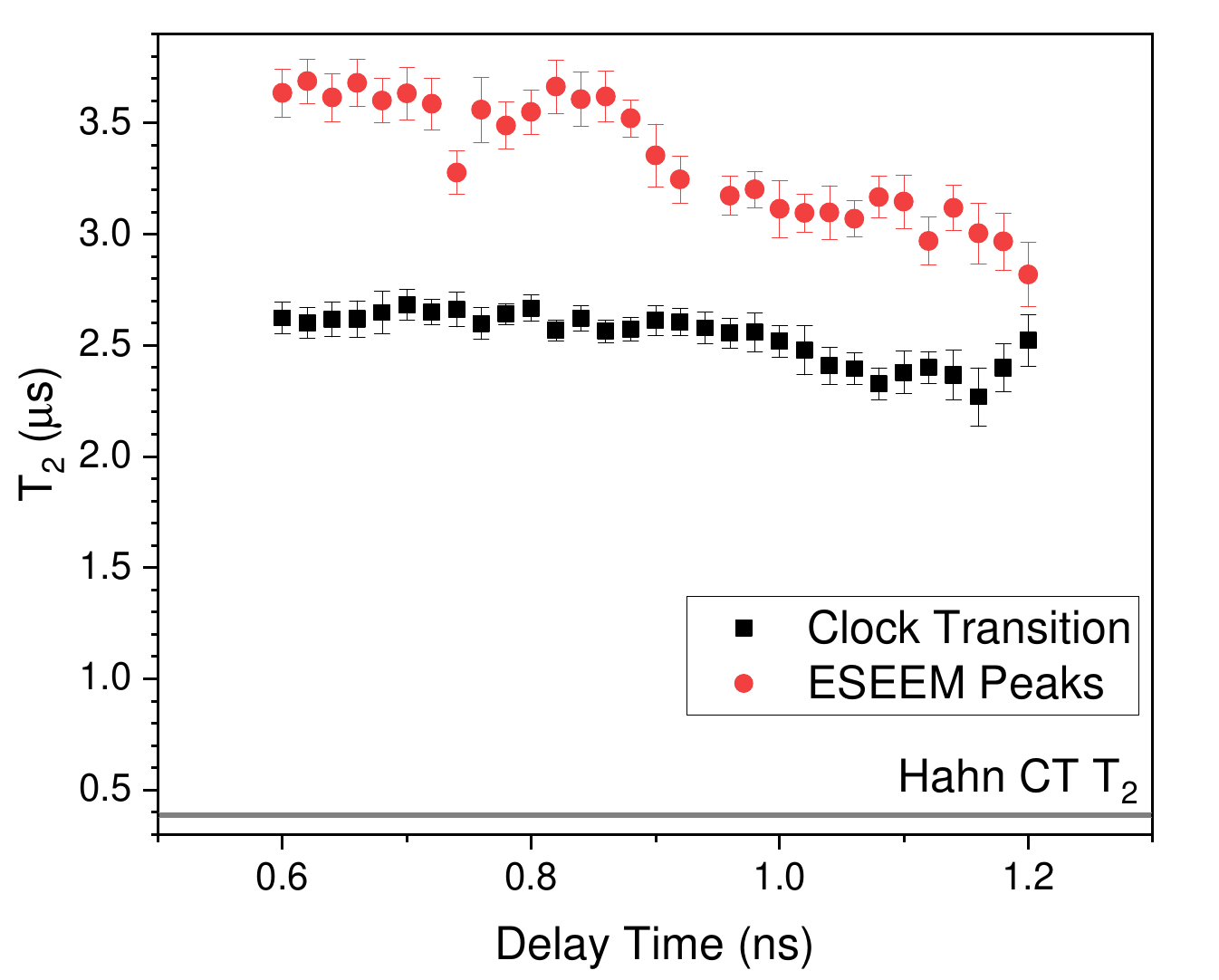}
    \caption{$\tau$ dependence of $T_2$ when using the CPMG sequence. $T_2$ at CT (black points) is $\sim2.6\,\upmu$s at small $\tau$ and falls off slightly for $\tau \gtrsim1~\upmu$s. The CPMG sequence was used to measure $T_2$ (red points) at the field for which $\tau$  suppressed the ESEEM oscillations -- see Fig.~\ref{fig:ESEEM_CPMG}. Measurements are from a $5$\% dilution sample of \textbf{2} in toluene at $5636$~MHz and $1.8$~K. The Hahn CT $T_2$ value indicated by a horizontal line near the bottom of the figure was measured along with the CPMG data. }
    \label{fig:CT_ESEEM_CPMG}
\end{figure}

Fig.~\ref{fig:CT_ESEEM_CPMG} shows $T_2$ as a function of the delay time $\tau$ obtained using the CPMG sequence. The effect of the CPMG sequence at the CT is shown in black. For small values of $\tau$, we obtain a $T_2$ value of $2.62(7)~\upmu$s. We observe that $T_2$ drops somewhat for $\tau \gtrsim1~\upmu$s, which confirms that CPMG works by filtering low-frequency decoherence~\cite{biercukDynamicalDecouplingSequence2011}, and gives some insight on the noise spectra in our system: that a significant portion of the noise is below $\sim\!1$~MHz.

\begin{figure}
    \centering
    \includegraphics[width=0.95\linewidth]{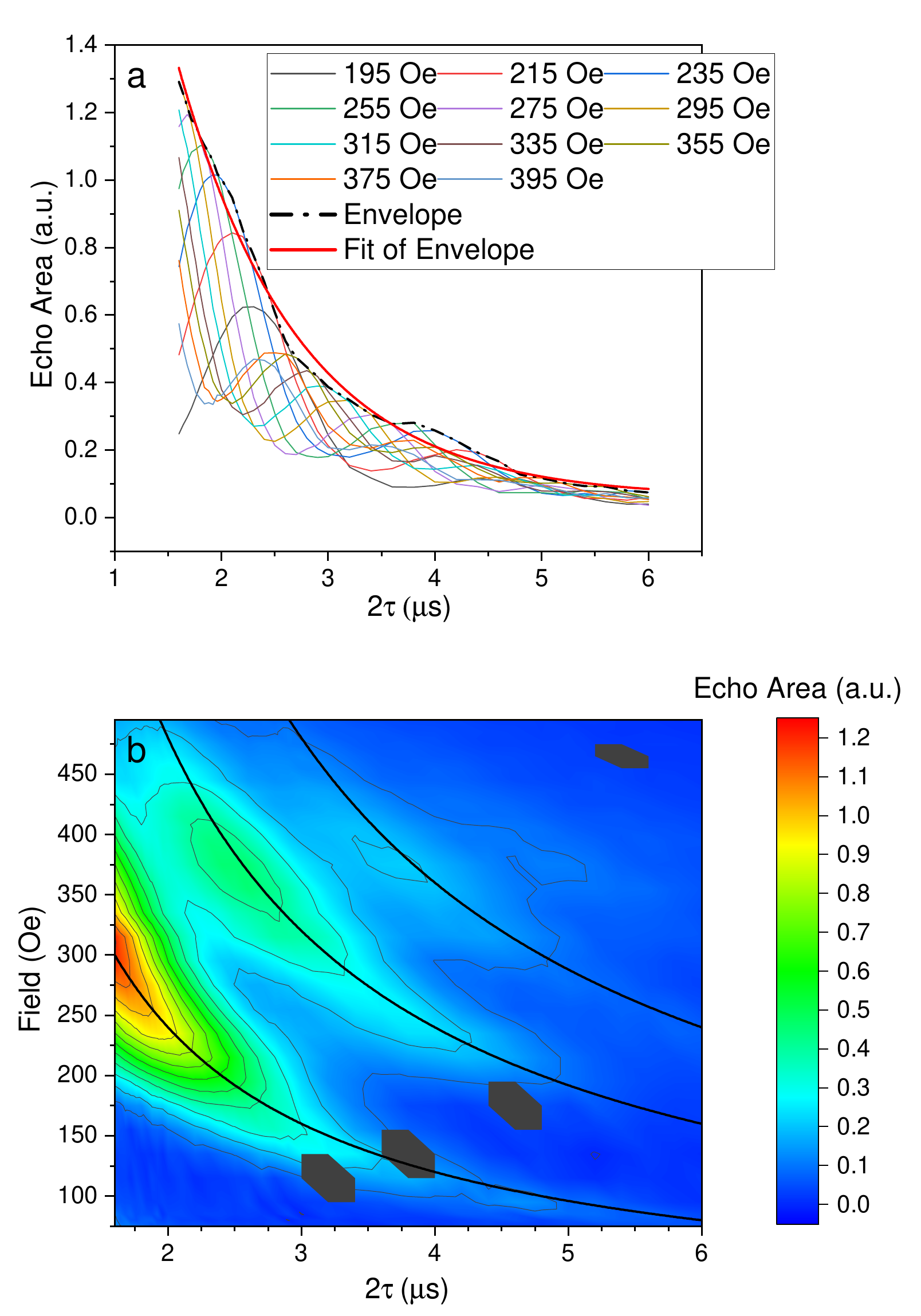}
    \caption{ESEEM as a function of delay time  and field. a) Echo area as a function of $2\tau$ at several field values, as indicated. The observed oscillations have frequencies that agree with the proton Larmor frequencies at the corresponding fields. The envelope of modulated data suggests that $T_2$ in this range of field does not change significantly and is comparable to the value around the CT. %
    b) Contour map of the echo area as a function of $2\tau$ and field value. Echo area is indicated by color. The dark black lines mark the loci where $\gamma_p B \tau = 2 \uppi n$, the expected position of the echo amplitude maxima, showing good agreement with data collected. Grey hexagons indicate places where data is missing due to control issues with the experimental apparatus. Measurements were taken on a $5$\% dilution sample of \textbf{2} in toluene at $1.9$~K and $4639$~MHz.}
    \label{fig:ESEEM_Osi}
\end{figure}

We next turn to the behavior near the second peak in Fig.~\ref{fig:EDFS}. Fig.~\ref{fig:ESEEM_Osi}(a) shows the echo area as a function of $2\tau$ for fields in the vicinity of that peak obtained using a Hahn sequence. The echo signal oscillates with $2\tau$ and the observed frequencies of oscillations matches well with the expected Larmor frequency of the proton at the corresponding field values, indicating the hyperfine origin of the oscillations. We collected oscillatory data at numerous field values and display the results in a contour map in Fig.~\ref{fig:ESEEM_Osi}(b), where echo amplitude, indicated by color, is plotted as a function of $2\tau$ and field. The oscillations arise from hyperfine interactions manifesting as ESEEM~\cite{jeschkeHyperfineDecouplingElectron1997}. The peak in echo signal occurs when $\omega_L \tau \approx 2 \uppi n$ and thus when $\tau \approx \frac{2 \uppi n}{\gamma_p B}$, using the Larmor frequency relation $\omega_L=\gamma_p B$, where $\gamma_p$ is the nuclear gyromagnetic ratio of the proton. These ESEEM oscillations are the origin of the second peak in Fig.~\ref{fig:EDFS}, which corresponds to the field for which $n=1$. The curves in the contour map show the expected relation between $2\tau$ and $B_{\text{peak}}$ for several integer values of $n$, indicating good agreement with the positions of the echo amplitude maxima. We note that while we assume that ESEEM is due to interactions with protons, consistent with what others have seen in similar heterometallic rings \cite{ardavanWillSpinRelaxationTimes2007}, the molecule also contains fluorine, which has a gyromagnetic ratio $\sim6\%$ smaller than that of the proton. With just a small number of ESEEM oscillations observable at low magnetic fields, we cannot rule out that fluorine plays a role in the observations.

\begin{figure}
    \centering
    \includegraphics[width=0.95\linewidth]{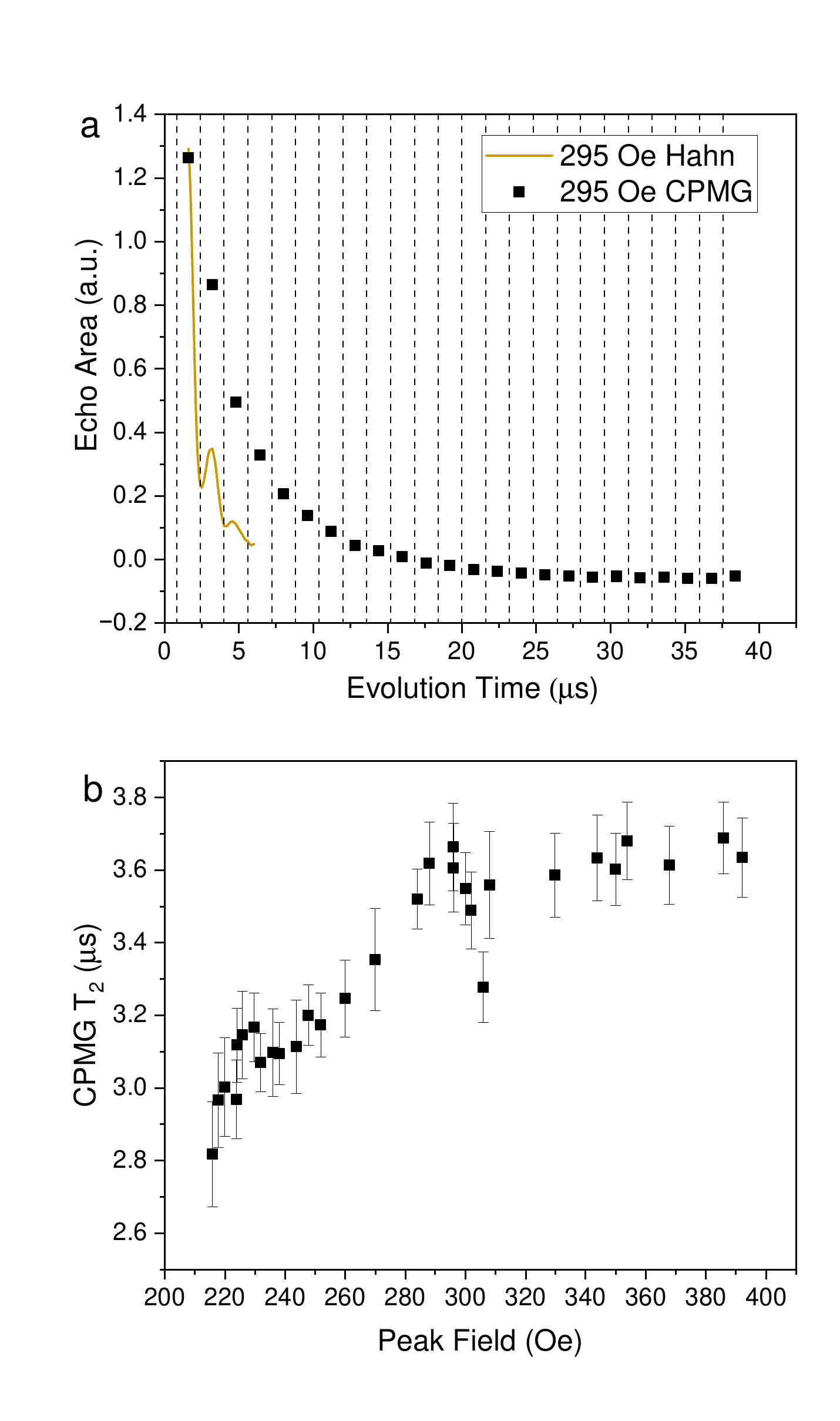}
    \caption{CPMG demodulation of ESEEM. a) Illustration of the decoupling technique at $295$~Oe. The yellow curve shows the Hahn-echo-measured ESEEM oscillations as a function of $\tau$. In our CPMG technique $\tau$ was set to match the Larmor period so that our $\uppi$ pulses are applied at the minima of the ESEEM oscillations in the Hahn data. The black points are the echo data obtained from this CPMG sequence; CPMG signal decays slower than that from the Hahn sequence. The dashed vertical lines indicate the times of the $\uppi$ pulses used in the CPMG sequence. The sample used for these measurements was a $5$\% dilution of \textbf{2} in toluene at 1.9~K and 4639~MHz. b) $T_2$ measured with the corresponding CPMG sequence at different field values with the technique illustrated in a). These measurements were carried out on a similar sample but at $1.8$~K and $5636$~MHz.}
    \label{fig:ESEEM_CPMG}
\end{figure}

Because of the oscillatory nature of the signal in Fig.~\ref{fig:ESEEM_Osi}, extracting a value of $T_2$ is difficult to do rigorously. In the Sec.~\ref{sec:coherent}, we present the results of simulations that allow us to fit the data to extract $T_2$. For now, we fit the envelope of the ESEEM to estimate $T_2$ at the ESEEM peaks as shown in Fig.~\ref{fig:ESEEM_Osi}(a). We find the ``coherence time'' around $1.13(5)$~$\upmu$s, which indicates in this field range the coherence is comparable to that found at the CT. %
Furthermore, we can employ the CPMG technique to dynamically decouple the electronic and nuclear spin dynamics and demonstrably enhance $T_2$~\cite{souzaRobustDynamicalDecoupling2012}. Fig.~\ref{fig:ESEEM_CPMG}(a) illustrates this technique at a single field of $295$~Oe. The yellow curve shows the Hahn-echo-measured ESEEM oscillations as a function of 2$\tau$. We measure CPMG with $\tau$ set to match the Larmor period -- so that our $\uppi$ pulses are applied at the minima of the ESEEM oscillations in the Hahn data. The resulting echo amplitudes, represented by the black squares in the figure, show that with this choice of $\tau$ the oscillations are no longer observable and the signal decays approximately exponentially. One can immediately see that the system's coherence has been extended through this technique since the echo amplitude obtained from CPMG is visible for a much longer time than when measured using the Hahn technique. A fit yields a value of $T_2$ of $3.96\left(1\right)\,\upmu$s for this sample.

We repeat this procedure for several fields using a similar sample, with the value of $\tau$ equal to the corresponding Larmor period. The value of $T_2$ extracted from each experiment is plotted as a function of field in Fig.~\ref{fig:ESEEM_CPMG}(b), showing an increase in $T_2$ from $\sim\!3.0$ to $\sim\!3.7~\upmu$s as field increases. The same data, plotted as a function of $\tau$ instead of field, is shown as red dots in Fig.~\ref{fig:CT_ESEEM_CPMG}. Notably, these values are all markedly larger than the values of $T_2$ obtained at the CT using the CPMG technique. Below we will present a similar conclusion from an analysis of our Hahn data. %

Summarizing the main findings in this section, we have shown that coherence in our Cr$_7$Mn system can be significantly enhanced at a zero-field CT and by an ESEEM process. This is despite the broad inhomogeneity in the sample, since limited energy diffusion allows us to coherently manipulate a small sub-ensemble. Using a period-matched CPMG pulse sequence, we dynamically decouple the electronic spin from the protons contributing to ESEEM, suppressing the oscillations and resulting in an enhanced coherence time. The observed behavior of this system near the CT is independent of variation in sample cation. Moreover, the protons in the solvent and intermolecular distance do not have much effect on coherence or ESEEM behavior, suggesting that decoherence processes in this system are dominated by sources within the molecule.

\section{Simulations -- Insights into decoherence from data}
To gain a better understanding of our data and of the underlying sources of decoherence affecting the system, we performed simulations of the dynamics. The simulations, along with fitting to the data, allowed us to extract information about decoherence both near the CT and in the ESEEM regime. Because of the prominent ESEEM in our data, simulations include coupling of the electronic spin with nuclear spins. In this work, we treat ESEEM through two approaches: coherent coupling of the electronic spin to a (set of) neighboring nuclear spin(s), and treatment of the nuclear spins as an environmental source of magnetic noise. While the former is approach is conceptually more straightforward, the latter is more expansive, allowing for the inclusion of incoherent noise sources that lead to decoherence. 

Standard treatments of ESEEM often involve a simple model of a $S=1/2$ electronic spin coupled to an $I=1/2$ nuclear spin. Such a model can be solved analytically~\cite{schweigerPrinciplesPulseElectron2001}. Here, we consider an $S=1$ system that has anisotropy and a CT. This complicates the coupling to nuclear spins and requires careful treatment. 

\subsection{ESEEM -- a coherent approach}
\label{sec:coherent}
A first approach to quantitatively treating our ESEEM results employs the following Hamiltonian
\begin{multline}
    \mathcal{H}=-D S_z^2+E (S_x^2-S_y^2)+g_s \mu_B\boldsymbol{B}\cdot\boldsymbol{S}\\
    - g_p\mu_p\boldsymbol{B}\cdot\boldsymbol{I} + A_{zz} S_z I_z + A_{zx}S_z I_x +A_{zy} S_z I_y,
    \label{eqn:ham_eseem}
\end{multline}
which in addition to Eq.~\ref{eqn:Cr7Mn_Ham} contains secular ($A_{zz}$) and pseudo-secular ($A_{zx}$, $A_{zy}$) hyperfine terms, as well as a nuclear Zeeman term. In calculating the dynamics of this model, we take into account the various forms of inhomogeneity of the system. 

As illustrated in Fig.~\ref{fig:levels}, there is substantial inhomogeneity in the transverse anisotropy $E$ of the spins. In addition, the use of solution samples in this study introduces orientational disorder to the system. Both the ESR transition frequency and transition matrix element depend on the molecule's orientation relative to the field. 
Thus, each sample contains a wide range of transition frequencies -- wider than the bandwidth of the resonator. This orientational disorder creates a complicated relationship between signal and molecular parameters that ultimately requires numerical modeling.

Despite the variety of orientations, the expected ESR signal in parallel mode is always along the direction of DC field. The direction of the field in the molecule's frame is along the direction $\hat{n}$ and we are interested in the component of spin along this direction, $S_n\equiv \boldsymbol{S}\cdot \hat{n}$.

For any given spin in our ensemble, we calculate the dynamics of the  quantum mechanical expectation value $\braket{\hat{S_n}}=\Tr\left(\rho \hat{S_n}\right)$. However, we also need to account for the various forms of inhomogeneity in the spin ensemble. We define the ensemble average $\overline{X}$ as the average of $X$ over the inhomogeneities in the sample, e.g.~random orientation of molecules as well as inhomogeneity in $E$. The ensemble average of $\braket{S_n}$ is thus %
\begin{align}
    \overline{\braket{S_n}} &=\frac{1}{4\uppi}\int\mathrm{d}\Omega \int \mathrm{d} E \lambda(E) L(\epsilon) \braket{S_n}\nonumber \\
    &=\frac{1}{4\uppi}\int\mathrm{d}\Omega \int \mathrm{d} \epsilon \left|\frac{\partial \epsilon}{\partial E}\right|^{-1} \lambda(E) L(\epsilon) \braket{S_n},
    \label{equ:ensemble}
\end{align}
where $L(\epsilon)=\left\{\left[2\left(\epsilon-\epsilon_0\right)/\Gamma\right]^2+1\right\}^{-1}$ is the lineshape function for the experimental resonator with $\epsilon$, $\epsilon_0$ and $\Gamma$ being the spin transition frequency, the resonator frequency and resonator linewidth, respectively. $\lambda(E)$ is the distribution of the transverse anisotropy parameter $E$, which we take to be sufficiently broad to be treated as a constant. The bounds of integration for $\epsilon$ are such that the entirety of the LGR's resonance is included, since only near-resonance signal is detected in the experiment. Finally, an integral over the solid angle $\mathrm{d}\Omega$ considers all possible orientations of the molecule in the solution. 

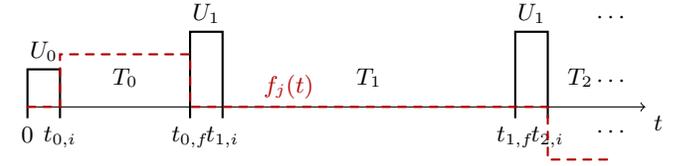
\begin{figure}
    \centering
    \begin{tikzpicture}[x=\linewidth/20, y=1cm]
    \tikzstyle{tick}=[thick]
    \tikzstyle{pulse}=[thick]
    \tikzstyle{label}=[font=\small]
    \tikzstyle{f} = [thick, dashed, red!70!black, line cap=round]
    \draw[->] (0,0) -- (19,0) node[below right, label] {$t$};
    \foreach \x in {0,1,5,6,15,16} {
        \draw[tick] (\x,0) -- (\x,-0.15);
    }
    \draw[pulse] (0,0) -- (0,.5) -- (1,.5) -- (1,0);
    \def\pulse#1#2{
        \draw[pulse] (#1,0) -- (#1,1) -- (#2,1) -- (#2,0);
    }
    \pulse{5}{6}
    \pulse{15}{16}
    \node[above, label] at (0.5,.5) {$U_0$};
    \node[above, label] at (5.5,1) {$U_1$};
    \node[above, label] at (15.5,1) {$U_1$};
    \node[above, label] at (18,1) {$\cdots$};
    \node[above, label] at (3,0.15) {$T_0$};
    \node[above, label] at (10.5,0.15) {$T_1$};
    \node[above, label] at (17,0.15) {$T_2$};
    \node[above, label] at (18,0.15) {$\cdots$};
    \node[below, label] at (0, -0.15)  {$0$};
    \node[below, label] at (1, -0.15)  {$t_{0,i}$};
    \node[below, label] at (5, -0.15)  {$t_{0,f}$};
    
    \node[below, label] at (6, -0.15)  {$t_{1,i}$};
    \node[below, label] at (15, -0.15) {$t_{1,f}$};

    \node[below, label] at (16, -0.15) {$t_{2,i}$};
    \node[below, label] at (18, -0.15) {$\cdots$};

    \draw[f] 
        (0,0) -- (1,0) -- (1, 0.7) -- (5, 0.7)
        -- (5,0) -- (6,0)
        -- (6, 0) -- (15, 0)
        -- (15,0) -- (16,0) -- (16, -0.7) --  (18,-0.7) ;
    \node[above right,red!70!black] at (7,0) {$f_j(t)$};
\end{tikzpicture}
    \caption{Schematic of CPMG-$N$ pulse sequence.  The notation used in simulation is illustrated and an example of $f_j(t)$ (see Sec.~\ref{sec:noise}) is shown. Note that since we treat pulses as infinitesimal, no noise-induced phase is accumulated during a pulse, hence $f_j(t)=0$ for all pulses.}
    \label{fig:sche}
\end{figure}
The dynamics of this model are implemented by assuming the system is initially in a psuedo-pure ground state, calculating the density matrix evolution $\rho\left(t\right)=U_t\rho\left(0\right)U_t^\dagger$ in the rotating frame of the radiation field, and then calculating $\braket{S_n}$. %
In the case of the CPMG sequence with $N$ $\uppi$ pulses (i.e., CPMG-$N$), the time evolution operator, $U_t=(U_\tau U_1U_\tau)^{N} U_0$, includes evolution due to the radiation pulses ($U_0$ and $U_1$ -- see Fig.~\ref{fig:sche}) as well as free evolution between pulses ($U_\tau$). The value of $\braket{S_n}$ is then averaged over the distribution in $E$ and in orientations (Eq.~\ref{equ:ensemble}) to produce a dependence of echo size on experimental variables like field and time. We add \textit{ad hoc} pure dephasing by multiplying the calculated time dependence with an exponential decay $e^{-t/T_2}$ with $T_2$ a free parameter that depends on field. Other parameters in our fitting are hyperfine coupling constants $A_{zz}$ and $A_{z\perp}\equiv A_{zx}$ (setting $ A_{zy}=0$, without loss of generality, due to symmetry), as well as an overall scaling factor. 

\begin{figure*}
    \centering
    \includegraphics[width=\linewidth]{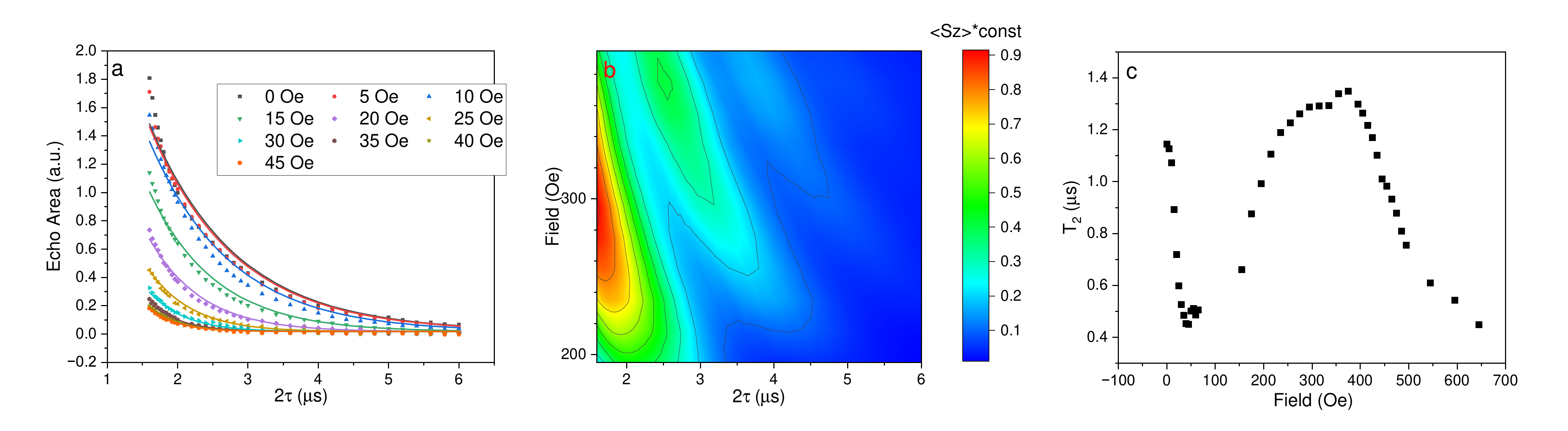}
    \caption{Fitting results in the fields around CT and 195 -- 395 Oe. In the simulation, we used infinitesimal pulses and free evolution with pure dephasing. The hyperfine parameters found with this fitting is $A_{zz}=5.72$~MHz, $A_{z\perp}=0.515$~MHz. a) Fitting (solid lines) of data (points) at fields near the CT. $T_2$ for each field value is treated as a free fitting parameter. 
    b) Contour map of the simulated signal as a function of $2\tau$ and field in range 195 -- 395~Oe. 
    c)  $T_2$ values extracted from fitting of data as a function of field.}
    \label{fig:fitting_t2}
\end{figure*}

This model works reasonably well in replicating our experimental data at fields near the CT and in the range $195$--$395$~Oe. Figure \ref{fig:fitting_t2}(a) shows a comparison of the experimental and simulated dependence for several fields near the CT. Unsurprisingly, in this regime no oscillations are observed since the Larmor frequency is very low. For fields in the range $195$--$395$~Oe, a 3D contour map of the simulation with the best-fit parameters is shown in Fig.~\ref{fig:fitting_t2}(b) and compares favorably with the data (Fig.~\ref{fig:ESEEM_Osi}) within the region of fields shown. 

Figure \ref{fig:fitting_t2}(c) shows the $T_2$ values determined from fitting at fields near the CT and at higher fields, excluding a range of fields where the fitting fails to match the data, discussed in more details below.
There are clearly two peaks in $T_2$: around the CT and in the vicinity of 300--400~Oe. Interestingly, we observe that the value of $T_2$ is somewhat larger at this second peak than at the CT (the smaller echo area is due to smaller matrix elements, i.e.~the undecohered signal is smaller). This is consistent with our finding that $T_2$ measured through CPMG is larger when measured in the ESEEM regime than at the CT (Fig.~\ref{fig:CT_ESEEM_CPMG}), but here we see this enhancement even without the dynamical decoupling provided by the CPMG sequence. This points to a difference in decoherence mechanisms at play in the two regimes: at zero field, the spin is immune to field fluctuations to first order and decoherence is driven by second-order fluctuations or fluctuations of a non-magnetic origin. At zero field, the nuclear spins have no Larmor precession and thus have incoherent dynamics. On the other hand, in the $\sim\!300$--$400$~Oe range, the nuclear spins have coherent Larmor precession, resulting in the observed ESEEM. This coherent precession may make the overall system dynamics more reversible, extending $T_2$ in this field range. 

Despite the success of this model near the CT and at higher fields, it fails to reproduce the observed behavior in the intermediate field range of $\sim\!50$--$200$~Oe -- see \cite{supp}. The signal size in this range of fields is approximately an order of magnitude smaller than in the higher-field range, something that we cannot adequately reproduce with the model presented above. %

In our fitting, we treated $T_2$ as a fitting parameter. However, $T_2$ is not a fundamental parameter, but a reflection of an underlying decoherence process. Such processes cannot generally be encapsulated by a single parameter. Thus, we turn to a more general approach to characterize how the environment gives rise to decoherence. In this approach, we characterize the environment as the source of a field-dependent noise spectrum seen by the electronic spin. 

To gain insight into this environmental noise and how it affects the dynamics of the electronic spin, we took detailed CPMG data with many values of $\tau$, a selection of which are shown in Fig.~\ref{fig:fittings}. (Additional data at other fields are provided in \cite{supp}.) Some recent studies have used such an approach with large numbers of $\uppi$ pulses to extract information about the noise spectrum of the environment~\cite{alvarezMeasuringSpectrumColored2011,bar-gillSuppressionSpinBathDynamics2012,myersDoubleQuantumSpinRelaxationLimits2017,williamsCharacterizingMagneticNoise2025,soetbeerRegularizedDynamicalDecoupling2021}. However, for the range of $\tau$ values accessible in our experiments, echoes from after more than three $\uppi$ pulses often had a poor signal-to-noise ratio, preventing us from implementing this approach directly. Instead, we used a physically motivated environmental noise model to simultaneously fit our CPMG-1 through CPMG-3 data, gleaning information about noise sources that underlie the observed behavior and decoherence.

\subsection{General approach to modeling decoherence}
\label{sec:noise}
We develop a phenomenological model of the noise as a function of field that includes a nuclear-spin bath, as well as noise in the molecule's transverse anisotropy. Using this model, we are able to extract information about the noise spectrum seen by the electronic spin that contributes to its decoherence. 

The complications arising from the system's inhomogeneities still require averaging using Eq.~\ref{equ:ensemble}. However, here we need an additional average to deal with the noise: $\E[\ldots]$ denotes the average of the decohering effects of noise. Including this average into our calculations changes our expected signal from $\overline{\braket{S_n}}$ to $\overline{\E[\braket{S_n}]}$.

To find $\E[\braket{S_n}]$, the decohered (noise ensemble averaged) expectation value of our signal along the DC field direction for a given value of $E$ and given orientation $\hat{n}$, we implement the following process. We treat our $S=1$ system as an effective two-level system, where the two lowest energy eigenstates, $\ket{\psi_0}$ and $\ket{\psi_1}$, are the pair involved in the CT at zero field and are resonant with the applied radiation. (At zero field, $\ket{\psi_0}=\ket{-}$ and $\ket{\psi_1}=\ket{+}$.)
We work in a rotating frame and employ the rotating-wave approximation so that the only time dependence in the Hamiltonian is produced by the noise during periods of free evolution. The dynamics of this system are described by the time-evolution operator: $\rho\left(t\right)=U\left(\delta\epsilon,t\right)\rho\left(0\right)U^\dagger\left(\delta\epsilon,t\right)$, where $\delta\epsilon\left(t\right)=\epsilon-\E[\epsilon]$ is the noise in transition energy $\epsilon$. We are interested in calculating $\braket{S_n}=\Tr\left(\rho\left(t\right) S_n\right)$ after time evolution. %
Any state in the two-level eigenbasis can be represented as $\Psi=\alpha \ket{\psi_0} + \beta \ket{\psi_1}$. It is then straightforward to show that $\braket{S_n}=2\,\Re\left( \alpha \beta^* \braket{\psi_1|S_n|\psi_0}\right)$. 

We wish to find the time evolution of the system after any number $N+1$ of radiation pulses (e.g.~CPMG-$N$ sequence). Following the same procedure discussed in coherence Sec.~\ref{sec:coherent}, the evolution is broken up into that produced by pulses and free evolution during inter-pulse time intervals, which we denote $T_m\equiv\left(t_{m,i},t_{m,f}\right)$, $m=0,1,\ldots$, as illustrated in Fig.~\ref{fig:sche}.
We treat the pulses as infinitesimally short, but not as perfect $\pi$ or $\pi/2$ pulses due to the inhomogeneity in $E$ and molecule orientation --  different spins in the ensemble will undergo different rotations. 
The inclusion of noise during inter-pulse intervals introduces additional complexity for a non-ideal sequence. In particular, the accumulated phase due to noise is mixed in a nontrivial manner.
For a given value of $E$, spin orientation, and noise history $\delta\epsilon\left(t\right)$, the dynamics of $\braket{S_n}$ reduces to 
\begin{equation}
    \braket{S_n}=\sum_j A_j \cos\phi_j,\label{jsum}
\end{equation}
where $A_j$ are constants that depend only on the pulses and $\phi_j$ is the accumulated noise-induced phase. $\phi_j=\sum_m a_{m,j} \int_{T_m} \mathrm{d}t\delta\epsilon=\int_T \mathrm{d}t\,\delta\epsilon f_j(t)$, defining $f_j(t)=\sum_{m} a_{m,j}\Theta(t-t_{m,i}) \Theta(t_{m,f}-t)$, where $\Theta(t)$ is the Heaviside step function, and $a_{m,j}$ is the weight of the phase accumulation for during the free-evolution time interval $T_m$. %
Fig.~\ref{fig:sche} shows an example of $f_j(t)$ in which $\{a_m\}_j=\{1,0,-1,\ldots\} $. With $N+1$ pulses, $m=0,1\ldots N$ for $N+1$ intervals. Each $a_{m,j}$ can be $0$ or $\pm 1$ for $m<N$. With these three possible values as well as the fact that $a_{N,j}=-1$ always (the accumulated phase in the last free evolution interval does not get mixed, as shown in example below), Eq.~\ref{jsum} has at most $3^{N}$ unique terms. 

As an example of the above procedure, we consider the two-pulse ``Hahn'' sequence, where the two periods of free evolution both have duration $\tau$. The evolution of the system initially in the ground state is described by the following.%
\begin{equation}
\begin{split}
    \begin{pmatrix}
        0 \\
        1
    \end{pmatrix} 
    & \xrightarrow[U_0]{}
     \begin{pmatrix}
        U_0^{00} & U_0^{01} \\
        U_0^{10} & U_0^{11}
    \end{pmatrix}\cdot
    \begin{pmatrix}
        0 \\
        1
    \end{pmatrix}=    
    \begin{pmatrix}
        U_0^{01} \\
        U_0^{11}
    \end{pmatrix}\\
    & \xrightarrow[U_\tau]{}
    \begin{pmatrix}
        U_0^{01} \\
        U_0^{11}e^{ i\int\mathrm{d}t_0\epsilon}
    \end{pmatrix}\\
    & \xrightarrow[U_1]{}
    \begin{pmatrix}
        U_1^{00}U_0^{01}+U_1^{01}U_0^{11}e^{ i\int\mathrm{d}t_0\epsilon} \\
        U_1^{10}U_0^{01}+U_1^{11}U_0^{11}e^{ i\int\mathrm{d}t_0\epsilon}
    \end{pmatrix}\\
    & \xrightarrow[U_\tau]{}
    \begin{pmatrix}
        U_1^{00}U_0^{01}+U_1^{01}U_0^{11}e^{ i\int\mathrm{d}t_0\epsilon} \\
        U_1^{10}U_0^{01}e^{ i\int\mathrm{d}t_1\epsilon}+U_1^{11}U_0^{11}e^{ i(\int\mathrm{d}t_0+\int\mathrm{d}t_1)\epsilon}
    \end{pmatrix}\\
    & \hspace{8em} \equiv \begin{pmatrix}
        \alpha \\
        \beta
    \end{pmatrix}
\end{split}
\end{equation}
For this case, we see $\alpha \beta^*$ (used to calculate $\braket{S_n}$) can be reduced to a sum of three terms (three values of $j$), corresponding to the three possible combinations of $\{a_0,a_1\}_j$, i.e.~$\{-1,-1\},\,\{1,-1\}$, and $\{0,-1\}$. After absorbing the phase due to $\E[\epsilon]$ into the $A_j$'s, we have three unique noise-induced phases $\phi_j$: $-\int\mathrm{d}t_0\delta\epsilon-\int\mathrm{d}t_1\delta\epsilon$, $\int\mathrm{d}t_0\delta\epsilon-\int\mathrm{d}t_1\delta\epsilon$, and $-\int\mathrm{d}t_1\delta\epsilon$.
As more pulses are added, e.g.~for a CPMG-$N$ sequence, each will increase the number of terms in Eq.~\ref{jsum}.%

Once $\braket{S_n}$ is determined (Eq.~\ref{jsum}), one can then calculate $\E[\braket{S_n}]$, assuming the noise has a Gaussian distribution with no temporal correlation, by using the well-known relation $\E[\cos \phi]=\exp(-\frac{1}{2}\E[\phi^2])$, yielding %
\begin{equation}
    \E[\braket{S_n}]= \sum_j A_je^{-\chi_j},
\end{equation}
where
\begin{multline}
    \chi_j=\frac{1}{2}\E[\phi_j^2]=\frac{1}{2}\iint_T\mathrm{d}t\mathrm{d}t' \E[\delta\epsilon(t) \delta\epsilon(t')]f_j(t)f_j(t')\\
    =\int_0^\infty \frac{\mathrm{d}\omega}{2\uppi} S_\epsilon(\omega) \left | \int_T \mathrm{d}t e^{ i\omega t}f_j(t)\right |^2.
\end{multline}
Here $S_\epsilon(\omega)\equiv \int\mathrm{d}t e^{i\omega t}\E[\delta\epsilon(0) \delta\epsilon(t)] $ is the spectral density of the noise $\delta\epsilon$. We define 
\begin{equation}
    \mathcal{F}_j=\omega^2\left| \int_T \mathrm{d}t e^{i\omega t}f_j(t)\right |^2,
\label{filter}
\end{equation}
which is commonly known as a filter function and is dependent on the particular pulse sequence described by $f_j$. CPMG (as well as other multi-pulse sequences) has a well-known filter function that can be used to analyze the dynamics of the experimental echoes to recover the underlying noise spectrum~\cite{cywinskiHowEnhanceDephasing2008,williamsCharacterizingMagneticNoise2025,malinowskiNotchFilteringNuclear2017,biercukDynamicalDecouplingSequence2011}. %

\subsection{Noise Sources -- Fitting the data}
We consider that the electronic spin sees two sources of noise:  one due to a nuclear spin bath and the other due to fluctuations in the anisotropy parameter $E$, which determines the CT frequency at zero field.  Most treatments of decoherence in MNMs focus on the effects of the spin bath~\cite{bar-gillSuppressionSpinBathDynamics2012,witzelConvertingRealQuantum2014,maClassicalNatureNuclear2015,zhangClusterCorrelationExpansion2020,schatzleSpinCoherenceStrongly2024,onizhukUnderstandingCentralSpin2024,ratiniMitigatingDecoherenceMolecular2025,joosProtectingQubitCoherence2022,malinowskiNotchFilteringNuclear2017}.  %
However, at the CT, decoherence from the spin bath is minimized, allowing other, non-magnetic sources of noise to dominate the decoherence.  We thus consider a form of noise that would remain if all magnetic fluctuations were absent: fluctuations in $E$, which could arise from 
dynamical conformational changes in the molecule. Given the large inhomogeneity in $E$ in this system, fluctuations in the parameter are unsurprising.  We assume these fluctuations in $E$ to have a $1/f$ spectrum. It is important to note that noise in $E$ cannot be filtered by the CT and so will always give rise to decoherence.  Importantly, we find that omitting the fluctuations in $E$ from our model results in a substantially worse fit to our zero-field data (see \cite{supp}).

Spin bath noise is modeled as arising from a collection of nuclear spins (protons with $I=\frac{1}{2}$), which we treat as isotropically distributed for simplicity. Each nuclear spin is taken to have a random orientation and to be uncorrelated with all other spins in the bath. It produces a dipole magnetic field at the location of the electronic spin that depends on its location in space. The total dipolar fluctuating field from the entire bath of nuclear spins at the electronic spin is dubbed $\boldsymbol{B}_I$. %
We treat $\boldsymbol{B}_I$ as a classical field. Thus, the spin Hamiltonian becomes
\begin{equation}
    \mathcal{H}=-D S_z^2+E (S_x^2-S_y^2)+g_s \mu_B(\boldsymbol{B}_{\text{ext}}+\boldsymbol{B}_I)\cdot\boldsymbol{S}.\\
    \label{eqn:ham_noise}
\end{equation}
The spectrum of the fluctuations in $\boldsymbol{B}_I$ is determined from the following simple model. Each nuclear spin has a random initial state and precesses at its nuclear Larmor frequency $\omega_L$ around the total instantaneous field $\boldsymbol{B}_{\text{ext}}$. %
We assume white-noise fluctuations in $\omega_L$ (reflecting interactions between bath spins). This, in turn, produces fluctuations in $\boldsymbol{B}_I$, with a Lorentzian spectrum with a central frequency $\omega_{L,0}$, corresponding to the mean value of field $B_{\text{ext}}$. Near the CT, $\epsilon$ has a non-linear dependence on field. Expanding this dependence in a Taylor series and keeping terms up to order $\boldsymbol{B}_I^2$ results in a noise spectrum in $\epsilon$ with peaks at $\omega_{L,0}$ and $2\omega_{L,0}$. %

Solving for the state energies from Eq.~(\ref{eqn:ham_noise}), we can infer how the temporal autocorrelation function of energy $\epsilon$ relates to correlation functions of $E$, as well as those for components of vector $\boldsymbol{B}_{I}$, and tensor $\boldsymbol{B}_{I}\otimes \boldsymbol{B}_{I}$. After averaging over the angular distribution of the spin bath and the random orientation of the nuclear spins, we obtain three ``conversion coefficients'' $C_i$, which are derived from the model before fitting, for any given specific anisotropy $E$ and $\boldsymbol{B}_{\text{ext}}$ (see \cite{supp} for details).
Combining these noise sources, one obtains a total noise spectral density of
\begin{equation}
    S_\epsilon(\omega)= C_1 S_{B_I}+C_2 S_{\boldsymbol{B}_{I}\otimes \boldsymbol{B}_{I}}+C_3 S_E,
\label{noiseS}
\end{equation}
where
\begin{align}
    S_{B_I}(\omega)&=\alpha\int\mathrm{d} t\,e^{i\omega t}   \cos\left(\omega_{L,0} t\right) e^{-\frac{1}{2}\sigma t},\\
    S_{\boldsymbol{B}_{I}\otimes \boldsymbol{B}_{I}}(\omega)&=\beta\int\mathrm{d} t\,e^{i\omega t}   \cos\left(2\omega_{L,0} t\right) e^{-{2}\sigma t},\\
    S_E&=\gamma\cdot\frac{1}{\omega}
\end{align}
where $\alpha$, $\beta$, $\gamma$ and $\sigma$ are parameters to be fit. $\alpha$ and $\beta$ contain information about the radial distribution of the spin bath, $\gamma$ represents the noise amplitude of fluctuations in $E$, and $\sigma$ represents the magnitude of the fluctuations in $\omega_L$. The fitting parameters $\alpha$ and $\sigma$ are taken to depend on the magnitude of $\boldsymbol{B}_{\text{ext}}$ while $\gamma$, $\beta$ and a scaling factor are treated as being independent of field.

\begin{figure}
    \centering
    \includegraphics[width=0.95\linewidth]{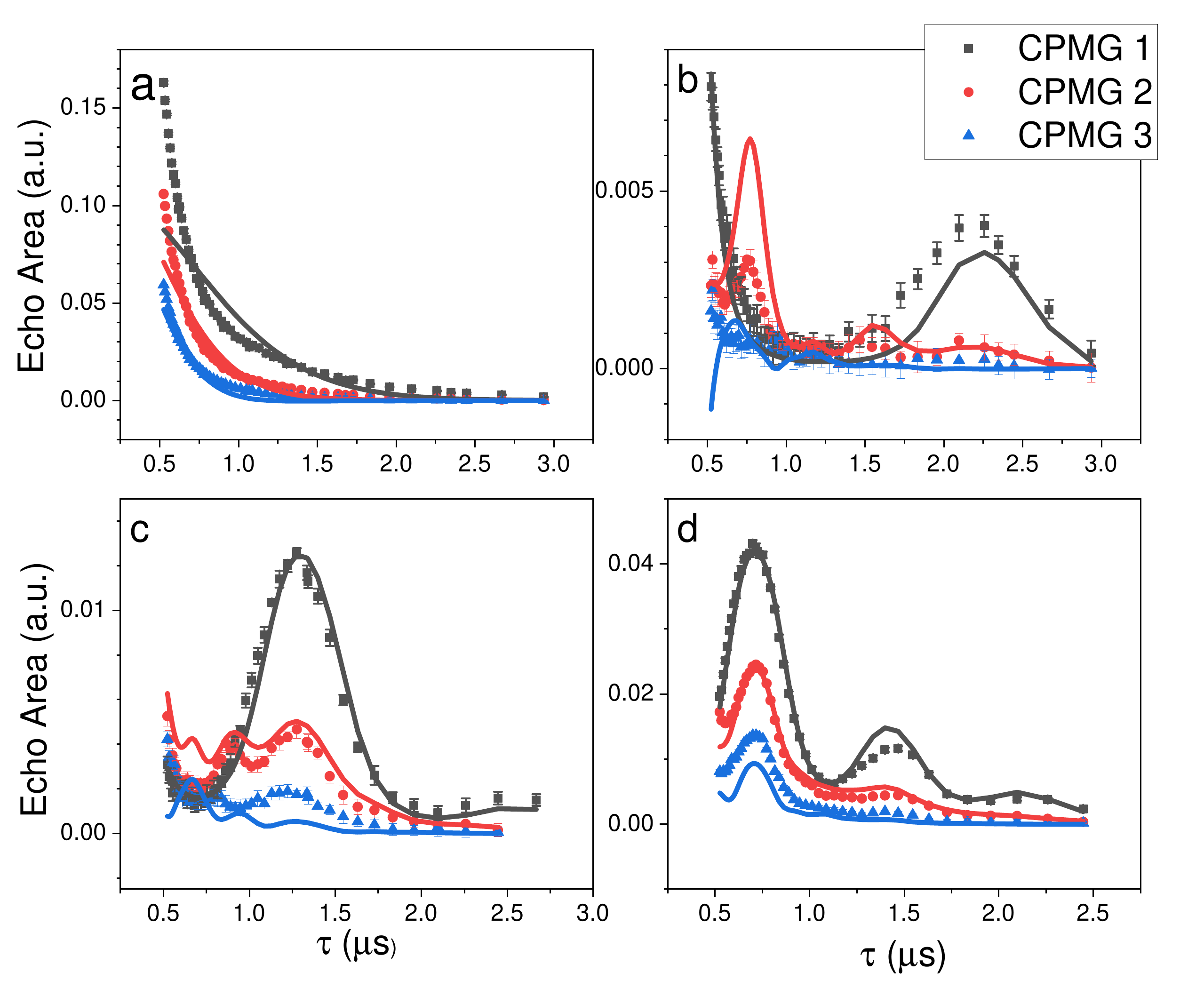}
    \caption{CPMG data (points) and fits (solid lines) at fields of (a) 0, (b) 100, (c) 175, and (d) 325~Oe. The sample is a $10$\% toluene solution of \textbf{2} at $1.8$~K and $4730$~MHz.}
    \label{fig:fittings}
\end{figure}

\begin{figure}
    \centering
    \includegraphics[width=0.95\linewidth]{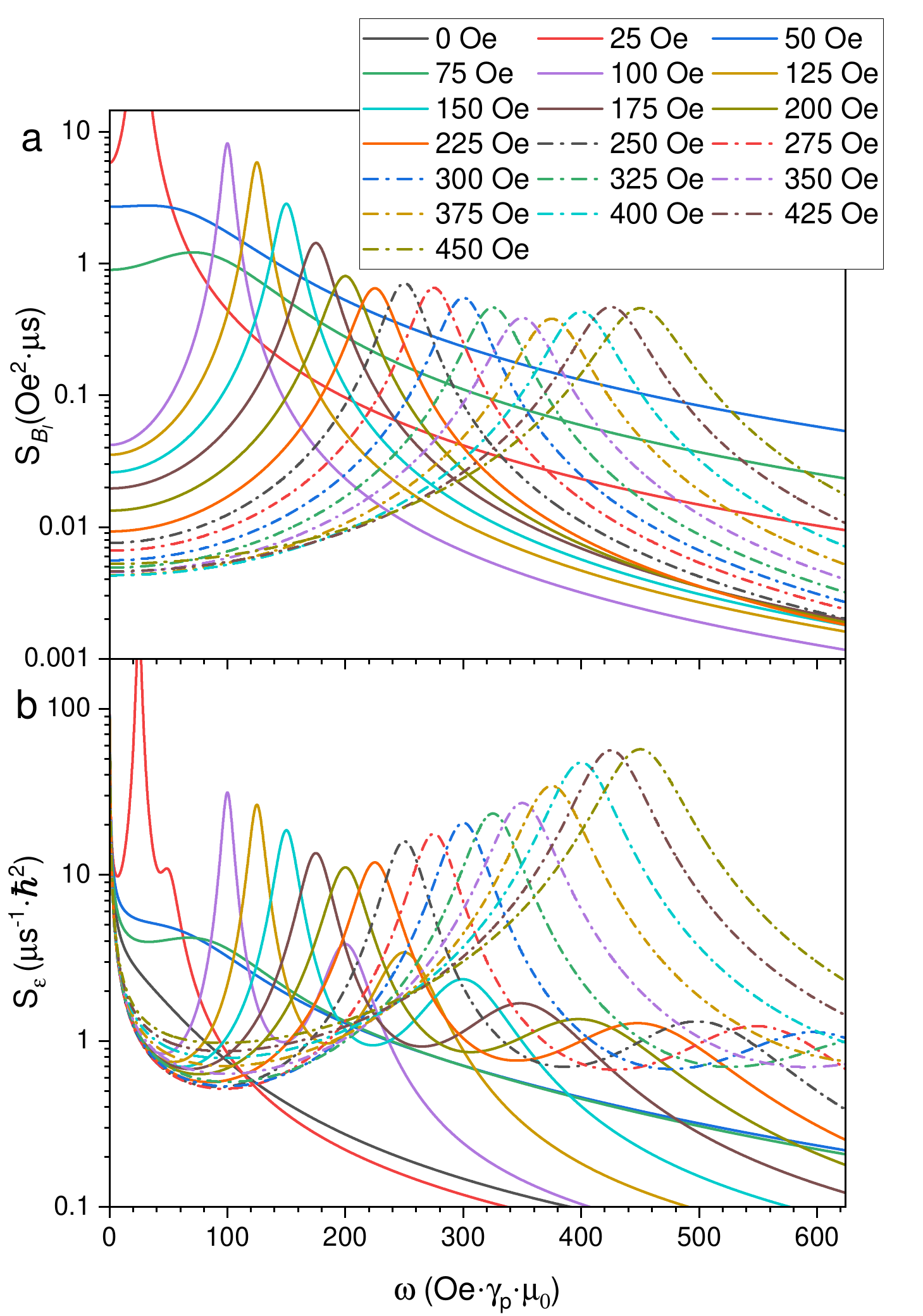}
    \caption{Noise spectra at different fields. a) $S_{B_I}\left(\omega\right)$ shows a single peak at the Larmor frequency. Note that no noise spectrum is shown for $B=0$ since it is effectively filtered out by the CT and therefore cannot be inferred from the experimental data. b) $S_\epsilon (\omega)$  determined from Eq.~\ref{noiseS}, calculated using the specific orientation that gives the largest contribution to the echo signal.}
    \label{fig:noise}
\end{figure}

Figure~\ref{fig:fittings} shows results of simultaneously fitting data from CPMG-1, CPMG-2 and CPMG-3, using all fields characterized in a single sample. For clarity, the figure shows results for fields of 0, 100, 175 and 325~Oe, as indicated. (Results of fitting at other fields are shown in \cite{supp}.)  There is good quantitative agreement at the two higher fields. The basic oscillatory structure is well reproduced at 100~Oe for both CPMG-1 and CPMG-2 even though the fit does not reproduce the data quantitatively. At certain fields, the simulations produce an additional peak (e.g.~$\tau\sim0.7\,\upmu$s in Fig.~\ref{fig:fittings}c) for CPMG-2 not seen in the data. 

We note that at zero field, the noise is dominated by fluctuations in the anisotropy $E$ since the CT effectively filters out field fluctuations. It may be surprising that the zero-field data is not reproduced well by the simulations since that behavior appears to be close to exponential. Indeed, one can fit, say, CPMG-1 alone and get a good fit, but the noise spectrum extracted will not reproduce the CPMG-2 or CPMG-3 behavior satisfactorily. This suggests that our model of treating the $E$ fluctuation spectrum as having a $1/f$ character does not fully capture the physics. We attempted other realistic models, such as a $1/f^2$ spectrum or adding white noise to the $1/f$ spectrum, without appreciable improvement.

Our fitting yields several parameters: one value of $\alpha$ and one value of $\sigma$ for each value of $B_{\text{ext}}$, and a single value for each of $\beta$, $\gamma$, and a scaling factor. From these parameters, we can calculate the components of the noise spectrum at each field (Eq.~\ref{noiseS}). Fig.~\ref{fig:noise}a shows the magnetic field component of the noise spectrum, $S_{B_I}$, for several values of $B$, as indicated. Notably, the amplitude of the Larmor peak (at $\omega_{L,0}$) is roughly constant for all fields above $\sim\!200$~Oe. 

Figure \ref{fig:noise}(b) shows the full noise spectrum $S_{\epsilon}$, containing all three components in Eq.~\ref{noiseS}. Since the conversion coefficients $C_1$ and $C_2$ depend on the orientation of the molecule with respect to the field direction, Fig.~\ref{fig:noise}(b) shows the spectrum that corresponds to the molecule orientation that yields the largest contribution to the echo signal for a given field. Other orientations yield similarly shaped spectra, but with different amplitudes for the Larmor and $1/f$ components. In the noise spectra shown, most fields show a prominent peak at $\omega_{L,0}$ (due to $S_{B_I}$), and a significantly smaller peak at $2\omega_{L,0}$ (due to $S_{\boldsymbol{B}_{I}\otimes \boldsymbol{B}_{I}}$), as well as the $1/f$ behavior at the lowest frequencies. %
In contrast to Fig.~\ref{fig:noise}(a), in $S_{\epsilon}$ we see that the overall area of the Larmor peak tends to steadily decrease with decreasing field (although for fields in the 100 -- 200~Oe range the width of this peak is substantially narrowed). This reflects the influence of the CT, where magnetic field noise is more effectively filtered out as the field approaches zero. The spectra at fields $<$100~Oe seem to vary widely with field, with $\sigma$ being anomalously large for 50 and 75~Oe. Unsurprisingly, it is in this same field range where the fits have only modest agreement with the data and so the inferred noise spectra may not fully represent the physics at those fields. Attempts to constrain $\sigma$ to values similar to those found at larger fields yielded substantially worse agreement with the data. 

\begin{figure}
    \centering
    \includegraphics[width=0.95\linewidth]{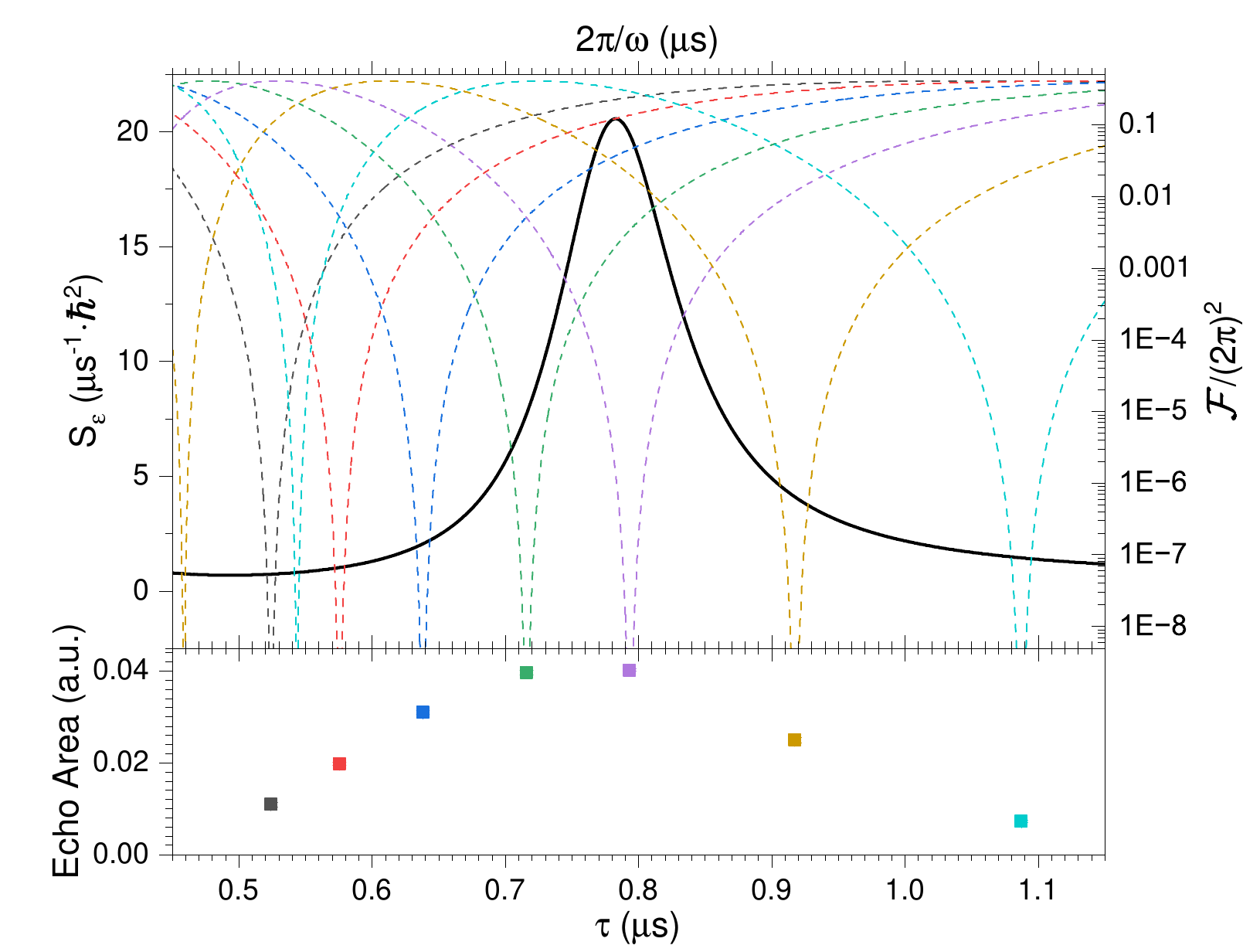}
    \caption{Illustration of ESEEM oscillations produced through filtering of noise. The black curve in the top panel is the noise spectrum at $300$~Oe as a function of $2\uppi/\omega$. The dashed lines indicate the Hahn (i.e.~CPMG-1) filter function $\mathcal{F}$ for different values of delay $\tau$; dips occur when $\omega =2n\uppi/\tau$. Values of $\tau$ chosen for the filter functions correspond to ones used in the experiment. When the dip of a filter function overlaps substantially with the peak in the noise spectrum, the noise is effectively filtered out. The lower panel shows measured echo area as a function of $\tau$ at $300$~Oe. When there is substantial noise filtering by the filter function, a significantly larger echo signal is observed.\label{fig:filter}}
\end{figure}

The observed behavior of our data as a function of field and delay time $\tau$ can be understood in terms of the filtering of the noise spectrum by our pulse sequence. Each pulse sequence has an associated filter function, $\mathcal{F}$, cf.~Eq.~\ref{filter}. For an ideal, two-pulse Hahn sequence (i.e.~CPMG-1), the filter function is given by $\mathcal{F}=16 \sin^4 \frac{\omega \tau}{2} $. In Fig.~\ref{fig:filter}, we overlay the noise spectrum $S_\epsilon$ at $300$~Oe with the filter function $\mathcal{F}$ for various values of $\tau$ used in our experiments. The filter function has a ``notch'' when $\omega =2\uppi/\tau$; noise at this frequency is filtered out very effectively. When the notch overlaps with the Larmor peak in $S_\epsilon$, the decoherence produced by that peak in the spectrum is significantly reduced, resulting in a large echo signal, as shown in the lower panel of the figure. As $\tau$ is varied, the effectiveness of this filtering is modulated, resulting in the oscillatory behavior characteristic of ESEEM. Note that this behavior does not rely on any intrinsic coherent coupling between the electronic spin and nuclear spins, but arises from simply treating the nuclear spins as the source of incoherent field-noise fluctuations peaked at the nuclear Larmor precession frequency. 

\section{Discussion}
There are a few notable conclusions we can attain from our results. One is that decoherence near the CT is primarily caused by noise sources arising within the molecule itself. This conclusion is supported by two findings: 1) We see very little change in behavior when we dilute the molecules in toluene solvent. Solutions ranging from 0.1\% to 10\% have been studied without appreciable change in the echo's observed dependency on field or in the $T_2$ values extracted. This implies that intermolecular interactions are not a significant form of decoherence. 2) Changing the solvent from regular toluene to deuterated toluene does not have any noticeable effect on our observations (see Fig.~\ref{fig:EDFS} and \cite{supp}). This suggests that the protons in the spin bath are within the molecule itself and not arising from the solvent matrix. This is consonant with previous work~\cite{wedgeChemicalEngineeringMolecular2012} on Cr$_7$Ni, which is a related ring but lacks a CT, that showed deuteration of solvent only had a substantial effect on coherence after the ligands had also been deuterated. The only form of long-range noise then that may give rise to decoherence would be spin-vibrational coupling, which is largely independent of dilution or of solvent deuteration.

Our  noise spectrum model has two distinct components: spin bath fluctuations and anisotropy fluctuations.  While the former may dominate for most spin qubits, the latter should be considered in systems that have CTs, where the magnetic fluctuations can be effectively filtered.  Indeed, fluctuations in the CT frequency itself may be the limiting source of decoherence for some CTs where the effects of the spin bath are minimized.  In our model, the zero-field CT frequency is determined entirely by $E$.  In other MNMs, the CT frequency is determined by other Hamiltonian parameters, such as elements of the hyperfine tensor, and the fluctuations in such parameters can lead to decoherence.  Such non-magnetic fluctuations are likely vibrationally mediated and thus may be mitigated through chemical engineering to alter the vibrational properties of the molecules, working at lower temperature to freeze out vibrational modes, or exploiting symmetries that suppress spin-phonon coupling of certain modes.

In addition, our extracted noise spectra (Fig.~\ref{fig:noise}) have strong dependence on field. While a zero-field CT is most effective in filtering field fluctuations when the external field is zero, that is also the condition for when the Larmor peak has moved to zero frequency. As we have shown, away from zero field, use of a period-matched CPMG sequence can effectively filter out the decohering effects of noise at the Larmor frequency. Such an approach cannot be employed at zero field where the nuclear spin dynamics are incoherent. Thus, the CT is the primary tool available for filtering both the noise from the spin bath and that of other magnetic degrees of freedom. If the noise from the nuclear spin bath could be substantially reduced by using, e.g., ligands with primarily $I=0$ isotopes, decoherence from Larmor precession would be mitigated. The CT could then potentially be more effective at filtering out the remaining magnetic field noise. 

The model we have developed of environmental noise, despite containing several heuristic and simplifying assumptions, demonstrates a remarkable ability to reproduce the key behaviors of the system. This suggests that the model captures the core mechanisms driving the system’s behavior. Since it is not very sensitive to specific microscopic configurations, the model has the potential to be applied to a broader family of systems (e.g., Cr$_7$Ni and other MNMs). It may serve as a flexible and extensible framework for understanding a wider range of related phenomena. 

The simplifying assumptions used in our model may account for discrepancies between the model and the actual behavior of the system: First, we assume an isotropic distribution of nuclear spins whereas the ring-like shape of the molecule suggests that the proton spins have an anisotropic distribution. Second, the Larmor precession frequency of the nuclear spins is taken to have some noise, modeled through the parameter $\sigma$, but the precession is taken to occur only around the applied field direction. At low applied fields, this assumption may not be well justified and the precession axis direction may itself be subject to noise. Next, our model is effectively one of ``pure dephasing,'' in which the noise only changes the relative phase in our two-state system, but we do not consider noise-induced transitions between energy eigenstates. %
Portions of our noise model, such as the $1/f$ character for $S_E$, have an \emph{ad hoc} character. Finally, we treat our noise as uncorrelated. A more realistic, many-body approach to treating the environment, particularly that of the spin bath, would be to include correlations between nuclear spins, using, e.g., a cluster-correlation expansion approach~\cite{onizhukColloquiumDecoherenceSolidstate2025,onizhukProbingCoherenceSolidState2021,baylissEnhancingSpinCoherence2022,ratiniMitigatingDecoherenceMolecular2025,zhangClusterCorrelationExpansion2020}. Such an approach may provide an understanding of the environmental noise that is better grounded in the microscopic structure of the molecule. 

We note that $T_2$ is found to be larger in the ESEEM regime than at the CT peak, when measured using both the Hahn  (Fig.~\ref{fig:fitting_t2}) and CPMG (Fig.~\ref{fig:CT_ESEEM_CPMG}) sequences. This perhaps surprising finding may be related to the nature of the nuclear spin dynamics: In the ESEEM regime, the nuclear spins have precession frequencies on the order of MHz. %
The contribution from this high-frequency field may be to add a modulation to the low-frequency noise, upconverting it to higher frequency, where it can be effectively filtered out by the pulse sequence. Modulation-induced upconversion of noise has been used as a technique to enhance coherence~\cite{joosProtectingQubitCoherence2022}. We suggest that something similar may be occurring in our system, albeit without an external modulation, but instead with the nuclear spin precession acting as a natural form of modulation. Thus, the enhancement of $T_2$ in the ESEEM regime may point to a built-in form of dynamical decoupling in the Cr$_7$Mn molecular nanomagnet.

Recent experimental studies have looked at CTs with frequencies well above X band~\cite{stewartEngineeringClockTransitions2024,gakiya-teruya546GHzClock2025}. %
Such high-frequency CTs have the potential for longer coherence times since the curvature of the frequency's dependence on field gets smaller with higher frequency. The lower frequency of the Cr$_7$Mn CT, in contrast, has some experimental and practical advantages. Experimentally, these frequencies are readily synthesized, allowing precise control of pulse sequences, including the phase needed for CPMG and other sequences. Practically, the CT studied here is similar in frequency to what is used in many superconducting qubits. In tandem, superconducting qubits' performance is degraded by large applied magnetic field.  Thus, the frequency range and low fields of MNM-based CTs may allow them to serve as elements in a hybrid quantum architecture \cite{clerkHybridQuantumSystems2020,gimenoOptimalCouplingHoW102023,jenkinsCouplingSinglemoleculeMagnets2013,carrettaPerspectiveScalingQuantum2021}, perhaps as part of a quantum memory, in which they are integrated with superconducting qubits. %
Thus, spin CTs with frequencies in the range of a few GHz are well suited for interfacing with superconducting qubits. MNMs like Cr$_7$Mn with low-frequency CTs are therefore attractive, potentially practical spin qubits. 

\section{Conclusion}
Our experiments reveal $T_2$ values in the few $\upmu$s range, when enhanced by CT or in a portion of the ESEEM oscillation field range.  Dynamical decoupling with the CPMG pulse sequence leads to enhanced coherence in both regimes.  
Studying molecular-spin qubits in the vicinity of CTs demonstrate the enhancement of coherence in these systems and provide a window into the underlying microscopic origins of decoherence.  In this study, through a combination of experimental data and theoretical modeling, we have found that decoherence in the Cr$_7$Mn molecular magnet is produced by nuclear-spin fluctuations, which can be substantially filtered by a CT, and fluctuations in the molecule's anisotropy, which cannot.  In systems with CTs, the fluctuations in the CT transition frequency, whether arising from fluctuations in anisotropy or from other Hamiltonian parameters, may present a limit on the ultimate efficacy of CTs in enhancing coherence.

One of the major advantages of MNMs is the ability to chemically engineer their properties.  Our findings about the decoherence sources in Cr$_7$Mn suggest ways to enhance coherence in this molecule and related MNMs.  Thus, we conclude that minimizing hyperfine fields from the nuclear spin bath -- even with the filtering offered by a CT -- is important to reduce their decohering effects.  In addition, stiffening the molecular structure or making other structural changes could reduce fluctuations in the CT frequency. Such design considerations should inform the development of new molecular-based spin qubits.  

\begin{acknowledgments}
We thank E.~Williams and N.~de Leon for useful conversations, K.~Ellers for early work on the experiment, A.~Anderson for assistance with use of the computing cluster, and J.~Kubasek and B.~Crepeau for significant technical assistance. This work was supported by the National Science Foundation under grant nos.~DMR-1708692 and DMR-2207624. JRF acknowledges the support of the Amherst College Senior Sabbatical Fellowship Program, which is funded in part by the H.~Axel Schupf ’57 Fund for Intellectual Life. This work was performed in part using high-performance computing equipment at Amherst College obtained under National Science Foundation Grant No.~2117377. Any opinions, findings, and conclusions or recommendations expressed in this publication are those of the authors and do not necessarily reflect the views of the National Science Foundation.

\end{acknowledgments}

\bibliography{Cr7Mnbib.bib}

\end{document}


\title{Supplementary Material for Enhancing Coherence with a Clock Transition and Dynamical Decoupling in the Cr$_7$Mn Molecular Nanomagnet}

\author{Guanchu Chen}%
\affiliation{Department of Physics and Astronomy, Amherst College, Amherst, MA 01002, USA}
\affiliation{Department of Physics, University of Massachusetts Amherst, Amherst, MA 01003, USA}%

\author{Brendan C.~Sheehan}%
\altaffiliation{Current address: Department of Physics and Astronomy, Dartmouth College, Hanover, NH 03755, USA}
\affiliation{Department of Physics and Astronomy, Amherst College, Amherst, MA 01002, USA}
\affiliation{Department of Physics, University of Massachusetts Amherst, Amherst, MA 01003, USA}%

\author{Ilija Nikolov}%
\altaffiliation{Current address: Department of Physics, Brown University, Providence, RI 02912, USA}
\affiliation{Department of Physics and Astronomy, Amherst College, Amherst, MA 01002, USA}

\author{James W.~Logan}%
\altaffiliation{Current address: Department of Physics and Astronomy, Dartmouth College, Hanover, NH 03755, USA}
\affiliation{Department of Physics and Astronomy, Amherst College, Amherst, MA 01002, USA}

\author{Charles A.~Collett}%
\altaffiliation{Current address: Department of Physics, Hamilton College, Clinton, NY 13323, USA}
\affiliation{Department of Physics and Astronomy, Amherst College, Amherst, MA 01002, USA}

\author{Gajadhar Joshi}%
\altaffiliation{Current address: Center for Integrated Nanotechnologies, Sandia National Laboratories, Albuquerque, NM 87123, USA}
\affiliation{Department of Physics and Astronomy, Amherst College, Amherst, MA 01002, USA}

\author{Grigore A.~Timco}
\affiliation{Department of Chemistry, The University of Manchester, Manchester M13 9PL, UK}

\author{Jillian E.~Denhardt}
\altaffiliation{Current address: University of Hawai`i at Mānoa, Honolulu, HI 96822, USA}
\affiliation{Department of Chemistry, University of Massachusetts, Amherst, MA 01003, USA}

\author{Kevin R.~Kittilstved}
\altaffiliation{Current address: Department of Chemistry, Washington State University, Pullman, 99164-4630, USA}
\affiliation{Department of Chemistry, University of Massachusetts, Amherst, MA 01003, USA}

\author{Richard E.~P.~Winpenny}
\affiliation{Department of Chemistry, The University of Manchester, Manchester M13 9PL, UK}

\author{Jonathan R.~Friedman}%
\email{jrfriedman@amherst.edu}
\affiliation{Department of Physics and Astronomy, Amherst College, Amherst, MA 01002, USA}
\affiliation{Department of Physics, University of Massachusetts Amherst, Amherst, MA 01003, USA}

\date{July 18, 2025}
\maketitle

\section{Inhomogeneous broadening}
\begin{figure}[hb]
    \centering
    \includegraphics[width=0.8\linewidth]{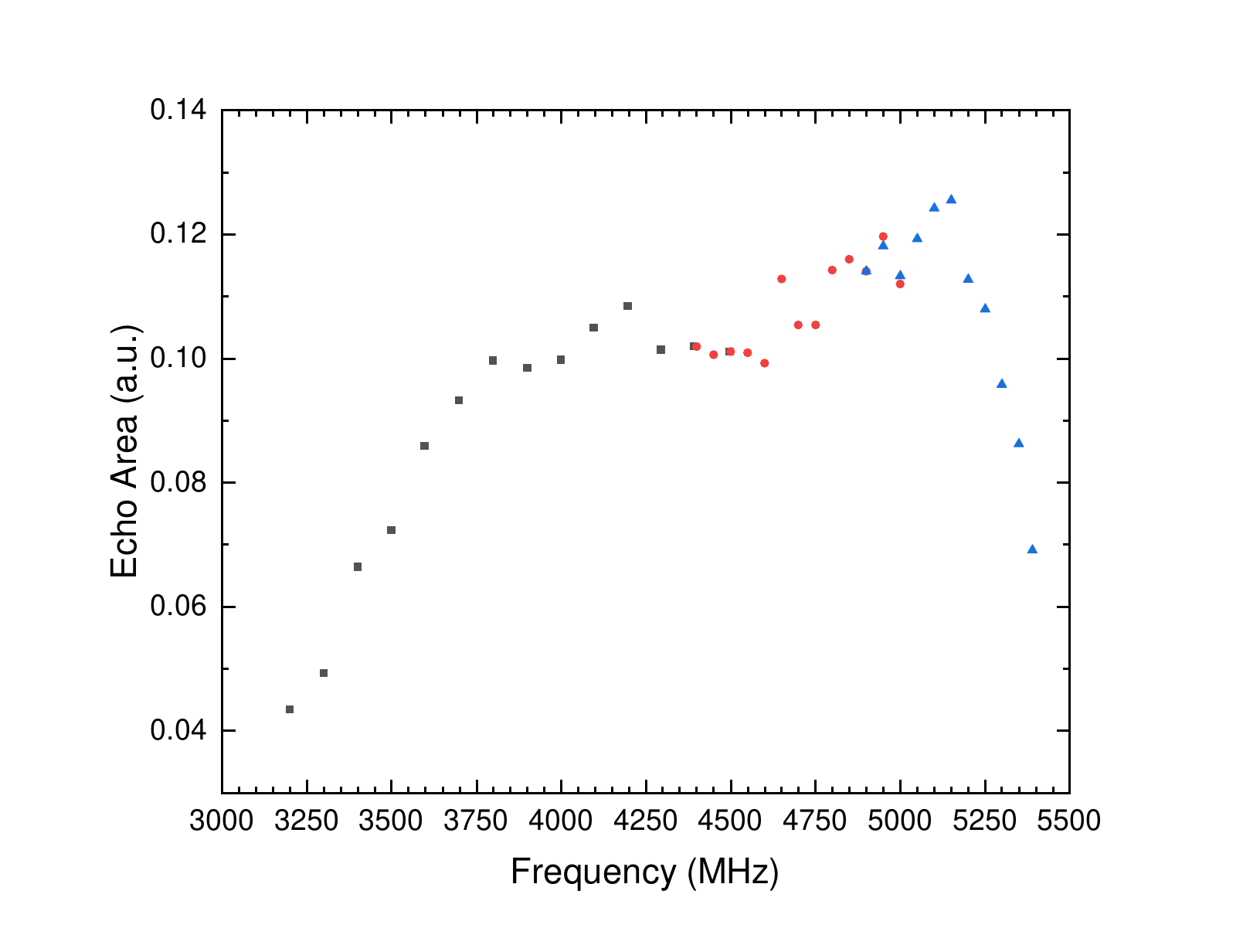}
    \caption{Zero-field frequency response of \textbf{2} at the clock transition. The data is from 1\% toluene solution sample at 1.8~K with three different resonators (marked with colors), indicating a very broad frequency response. Since this is a zero-field signal, we can infer a broad inhomogeneity in the anisotropy parameter $E$.
    }
    \label{fig:freq_echo}
\end{figure}

The inhomogeneous broadening can be seen in the broad response of ESR echo signal over a wide range of frequencies. As shown in Fig.~\ref{fig:freq_echo}, at zero field we see significant echo signal across a frequency range of $\sim\!2.5$~GHz, with 3 different resonators -- the frequency of each resonator is tuned over the range indicated by a single color through use of a dielectric. At zero field, the transitions frequency is determined exclusively by the transverse anisotropy:  $\epsilon=2E$;  thus, the observations provide a direct indication of the inhomogeneity in $E$. It is worth noting that we have seen echo beyond this range; the data shown is from a particular experiment with controlled frequency response.

\section{hole burning}
\begin{figure}[ht]
    \centering
    \includegraphics[width=\linewidth]{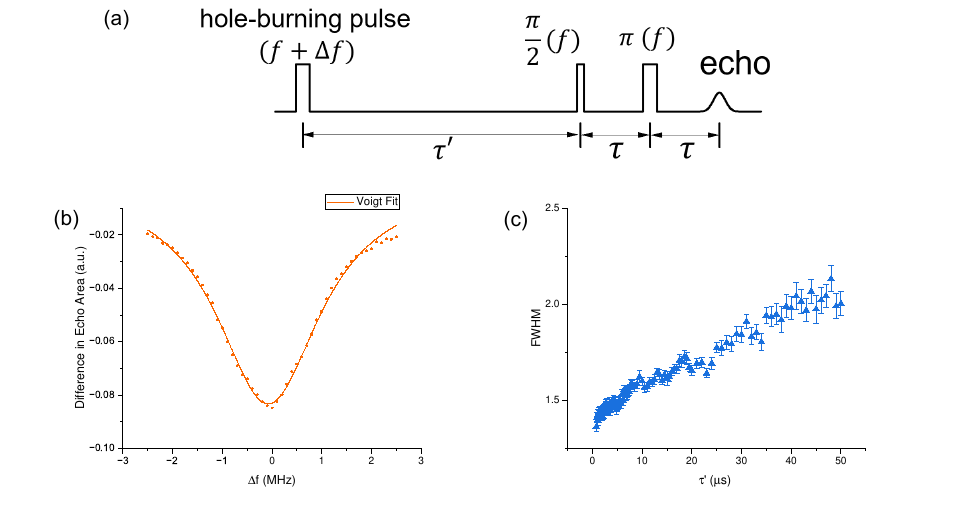}
    \caption{Hole Burning Experiment. %
    a) Pulse diagram of the hole-burning  sequence. A delay of $\tau'$ follows the  
    hole-burning pulse after which echo is detected with a Hahn sequence. b) Example of a burned hole. The hole-burning pulse is ``soft'', having a long width of $2~\upmu$s. Data is measured with $10$\% dilution \textbf{2} toluene solution at $1.8$~K and $5625$~MHz. c) Evolution of FHWM of the burned hole. The diffusion happens slowly compared to the typical timescale of our pulsed experiments, i.e.~$T_2$: after $50~\upmu$s, the hole width increases from $\sim\!1.5$~MHz to  $\sim\!2$~MHz.
    }
    \label{fig:holeBurning}
\end{figure}

Despite the broad inhomogeneity, at any given drive frequency, we are only exciting a small sub-ensemble of the spins.  This is demonstrated with a hole-burning experiment, illustrated in Fig.~\ref{fig:holeBurning}(a). The sequence contains three pulses: a hole-burning pulse is followed after a time $\tau'$ by a traditional Hahn echo sequence ($\uppi/2$-$\tau$-$\uppi$-$\tau$-echo), where $\tau$ is the delay time between Hahn pulses. The hole-burning pulse is detuned from the working ESR frequency by some detuning frequency $\Delta f$, which becomes the independent variable of the experiment. The hole-burning pulse has a duration of $2~\upmu$s so that its linewidth is small compared to those of Hahn sequence pulses. The amplitude of the hole-burning pulse is kept small to prevent possible saturation effects. By sweeping $\Delta f$ a hole is measured (an example is shown in  Fig.~\ref{fig:holeBurning}(b)) and fit with a Voigt function. By changing time $\tau'$, we are able to map the energy diffusion over time. From  Fig.~\ref{fig:holeBurning}(c), we see that the full width at half maximum of the hole changes from $\sim\!1.5$~MHz to $\sim\!2$~MHz over $50~\upmu$s of evolution. Our typical pulsed experiments were done within the time scale of tens of microseconds, meaning we only probed the resonant sub-ensemble of the system, before any significant spectral diffusion had occurred.

\section{Comparison of Sample variants, dilutions and cooling techniques}
\begin{figure}[htb]
    \centering
    \includegraphics[width=0.8\linewidth]{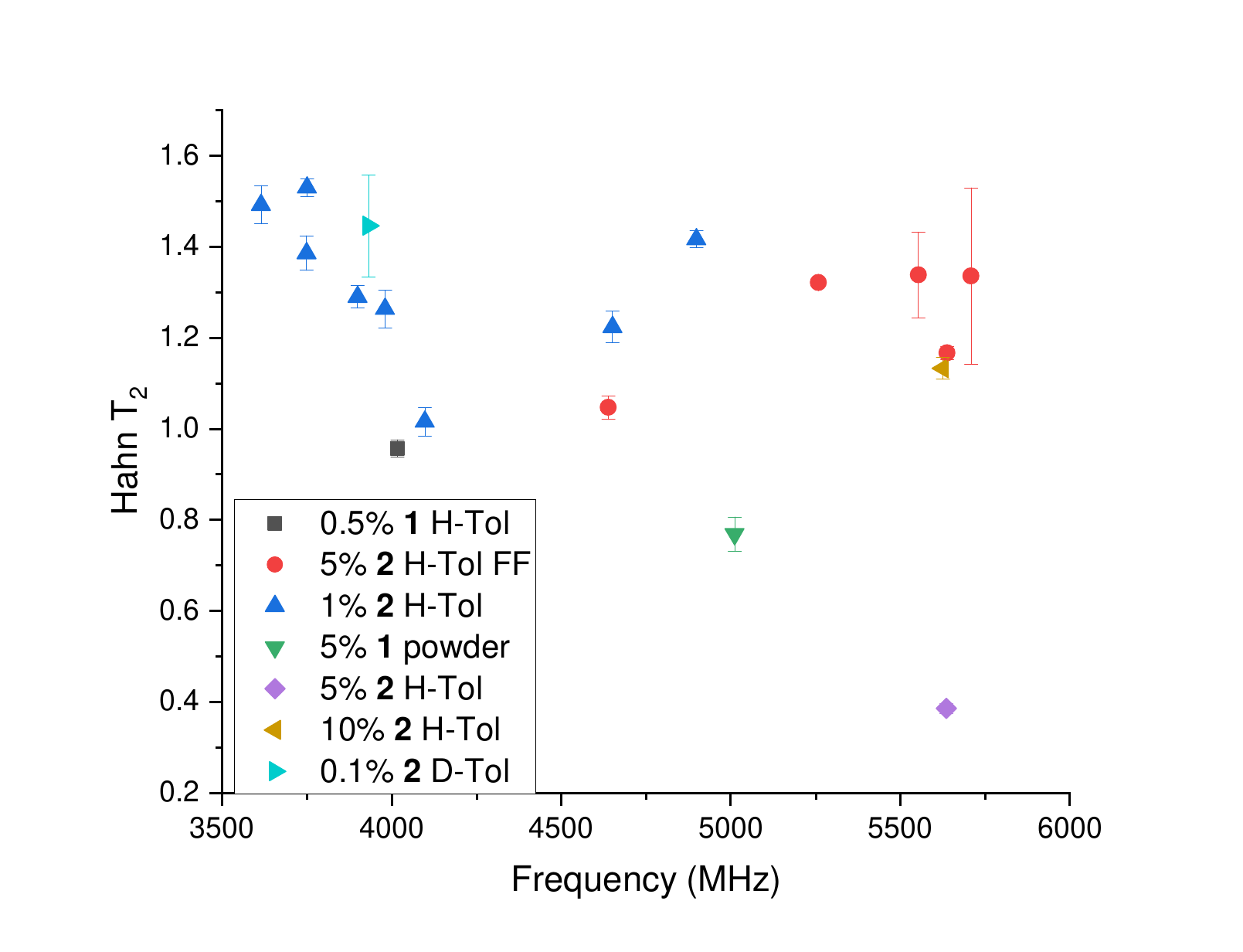}
    \caption{Frequency dependency of Hahn $T_2$ at the zero-field clock transition from different sample variants, dilutions, solvent and cooling techniques.  D-Tol indicates deuterated toluene,  FF indicates a sample that was flash frozen, and powder refers to a solid solution sample of Cr$_7$Mn in Ga$_7$Zn. 
    }
    \label{fig:freq_samples}
\end{figure}

\begin{figure}[htb]
    \centering
    \includegraphics[width=0.8\linewidth]{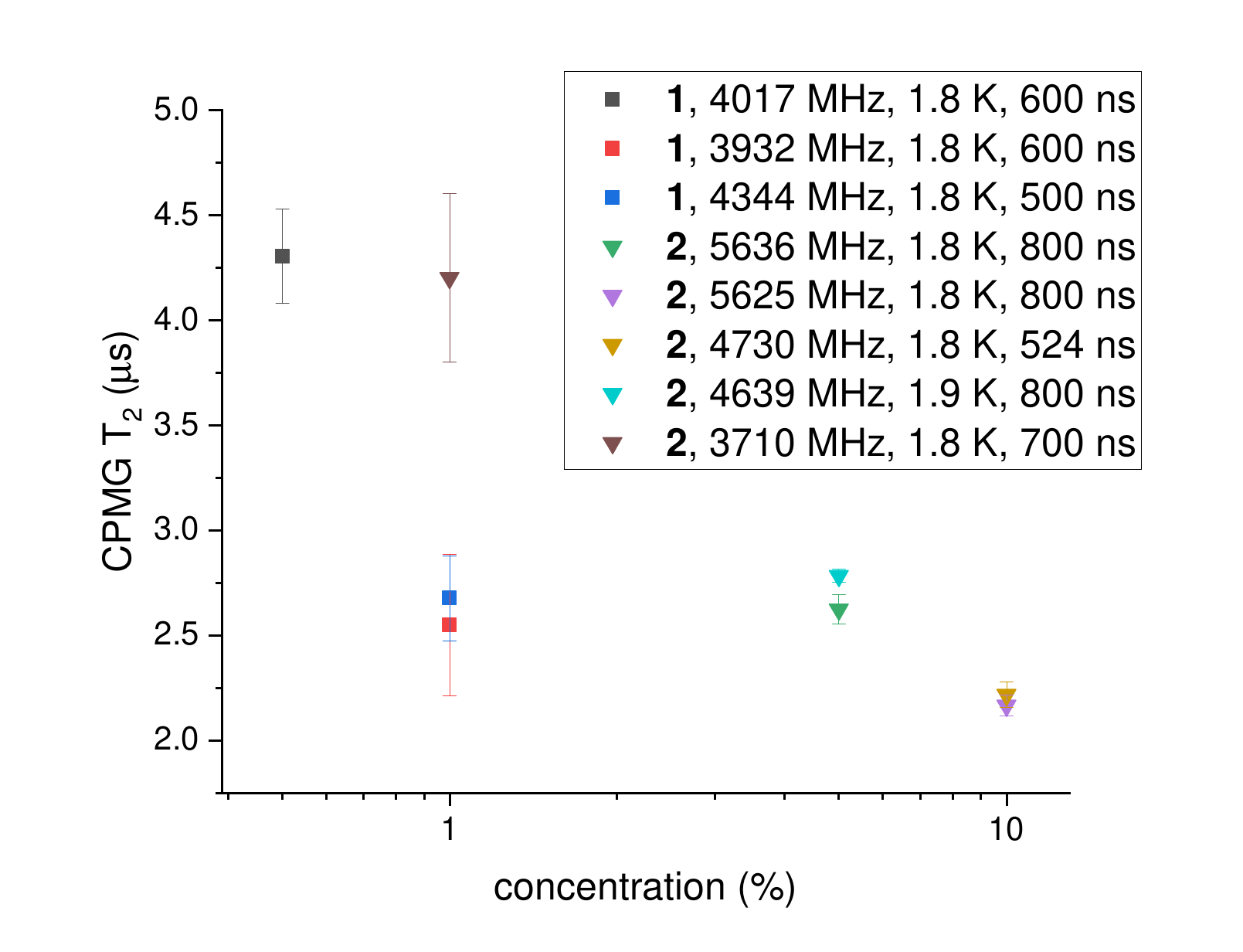}
    \caption{CPMG $T_2$ dependence on dilution. The data is from various samples, as noted in legend, with frequencies, temperature, and $\tau$ of the CPMG sequences. The data appear to indicate longer $T_2$ with lower dilution, but the scatter makes it hard to infer strong conclusions. }
    \label{fig:cpmg_samples}
\end{figure}

Fig.~\ref{fig:freq_samples} shows a comparison of echo signals from different Cr$_7$Mn sample variants, dilutions and cooling techniques. While signal strength naturally varies -- largely due to concentration and also due to variations in the measurement circuit and other experimental conditions -- we expect the coherence time $T_2$ to be more governed by intrinsic physical properties, and to exhibit more consistent and interpretable trends. The figure compares $T_2$ values for several samples, measured at a variety of frequencies.  The most naive expectation is that $T_2$ should be enhanced by further dilution.  However, we see very little change as concentration varies between 0.1\% and 10\%.   When using the CPMG sequence -- Fig.~\ref{fig:cpmg_samples} -- we do see some hint of a dependence of $T_2$ on concentration, but it is hard to draw strong conclusions from the data. %

The measured $T_2$ value also does not seem to depend significantly on the solvent: liquid solutions using toluene or deuterated toluene solvents were studied as well as solid solutions by co-crystallizing Cr$_7$Mn with Ga$_7$Zn; all show rather similar behavior.  Variation of cation (comparing \textbf{1} with \textbf{2}) was also not found to have significant effect on the coherence. %
This indicates that the most pronounced noise contributing to $T_2$ originates from within the molecule itself.

\section{Noise Spectrum Conversion Coefficients}
We derive the conversion coefficients from the noise on $\boldsymbol{B}$ to noise on $\epsilon$ by solving the Hamiltonian and then expanding it around the external field $\boldsymbol{B}_0$. In Einstein notation, denoting $\epsilon_i \equiv \frac{\partial\epsilon}{\partial B^i}$ and $\epsilon_{ij} \equiv \frac{\partial^2\epsilon}{\partial B^i\partial B^j}$, a Taylor series expansion yields
\begin{equation}
    \epsilon = \epsilon_0 +  \underbrace{B^i\epsilon_i}_{2nd} +  \underbrace{B^i B^j\epsilon_{ij}}_{3rd} + \underbrace{\frac{\partial \epsilon}{\partial E} \delta E}_{4th},
    \label{equ:taylor}
\end{equation}
where $B^i\equiv \int r^2 \mathrm{d}r \mathrm{d}\Omega  b^i (\boldsymbol{r})$ is the collective induced field of the proton bath at the origin (location of the electronic spin), and $\delta E$ is the dynamical deviation of anisotropy parameter $E$. The field $b^i (\boldsymbol{r})$ is defined below, cf.~Eq.~\ref{eq:b}. It is easy to see there is no cross-correlation between terms (i.e., 2nd, 3rd, and 4th terms). Thus, to get the temporal autocorrelation of $\delta\epsilon=\epsilon-\E[\epsilon]$, one can calculate the contribution of each term separately. For the second term in Eq.~\ref{equ:taylor}, we can get the contribution to $\E[\delta\epsilon(t_1)\delta\epsilon(t_2)]$:
\begin{equation}
\corr_2\equiv\E[\delta\epsilon(t_1)\delta\epsilon(t_2)]=\epsilon_i \epsilon_j \E[ B^i(t_1)  B^{j}(t_2)].
\end{equation}
We assume the field coming from protons at different locations is uncorrelated, so
\begin{equation}
\E[ B^i (t_1)  B^{j} (t_2)]=\int r^2 \mathrm{d}r \mathrm{d}\Omega \E[ b^i (\boldsymbol{r},t_1) b^{j}(\boldsymbol{r},t_2)]
\end{equation}
As shown in the next section, $\E[ b^i (\boldsymbol{r},t_1) b^{j}(\boldsymbol{r},t_2)] = R_2(r)g^{i,j}(\theta,\phi) \cos(\omega_{L,0}T)\exp(-\frac{1}{2}\sigma T)$ with $T\equiv t_2-t_1$, given our assumption that the noise arises from Larmor precession at frequency $\omega_{L,0}$.
Then
\begin{align}
    \corr_2 &= \cos(\omega_{L,0} T)e^{-\frac{1}{2}\sigma T} \int R_2(r) r^2 \mathrm{d}r \int \mathrm{d}\Omega  g^{i,j}(\theta,\phi) \epsilon_i\epsilon_j 
    \label{equ:corr2}\\
    & \equiv \alpha C_1 \cos(\omega_{L,0} T)e^{-\frac{1}{2}\sigma T},
\end{align}
defining $\alpha \equiv \int R_2(r) r^2 \mathrm{d}r$ a fitting parameter reflecting the radial distribution in protons, and $C_1\equiv \epsilon_i\epsilon_j\int \mathrm{d}\Omega  g^{i,j}(\theta,\phi)$, which can be calculated assuming an isotropic distribution of protons.

The same process can be carried out for the 3rd and 4th terms to obtain
\begin{align}
    \corr_3 &=\cos(2 \omega_{L,0} T)e^{-2\sigma T} \int R_4(r) r^2 \mathrm{d}r \int \mathrm{d}\Omega  g_2^{ijkl}(\theta,\phi) \epsilon_{ij}\epsilon_{kl}\label{equ:corr3}\\
    &\equiv \beta C_2 \cos(2 \omega_{L,0} T)e^{-2\sigma T},\\
    \corr_4 &=  \left(\frac{\partial \epsilon}{\partial E}\right)^2 \E[\delta E (t_1)\delta E (t_2)] \\
    &\equiv C_3\E[\delta E (t_1)\delta E (t_2)],
\end{align}
in which $\beta \equiv \int R_4(r) r^2 \mathrm{d}r$, $C_2\equiv\epsilon_{ij}\epsilon_{kl}\int \mathrm{d}\Omega  g_2^{ijkl}(\theta,\phi)$, $C_3\equiv \left(\frac{\partial \epsilon}{\partial E}\right)^2$, $\E[\delta E (t_1)\delta E (t_2)] = \int \frac{\mathrm{d}\omega}{2\uppi} e^{-i\omega t} \gamma \cdot \frac{1}{\omega}$, following the \emph{ad hoc} assumption of a $1/f$ spectrum for the noise in $E$.

\section{Noise on Larmor frequency}
We treat each nuclear spin classically, as a  magnetic dipole moment $\boldsymbol\mu$ at position $\boldsymbol r$. The induced magnetic field at the electronic spin located at the origin is
\begin{align}
b^i (\boldsymbol{r})& =\frac{\mu_0\mu}{4\uppi r^3} \left( 3\hat{r}^i\hat{r}_m\hat{\mu}^m-\hat{\mu}^i\right)\label{eq:b}\\
&=\frac{\mu_0\mu}{4\uppi r^3} \left( 3\hat{r}^i\hat{r}_m-\delta^i_m\right) \hat{\mu}^m
\end{align}

Then, for $\corr_2$, one obtains
\begin{align}
\E[ b^i(\boldsymbol{r},t_1)  b^{j}(\boldsymbol{r},t_2)]
&=R_2(r) \left( 3\hat{r}^i\hat{r}_m-\delta^i_m\right)\left( 3\hat{r}^j\hat{r}_n-\delta^j_n\right)\E[\hat{\mu}^m(t_1)\hat{\mu}^n(t_2)]\\
&\equiv R_2(r) \Theta^{ij}_{mn} (\theta,\phi) \E[\hat{\mu}^m(t_1)\hat{\mu}^n(t_2)],
\label{equ:corr2b}
\end{align}
where  $R_2(r)=\left(\frac{\mu_0\mu}{4\uppi r^3}\right)^2$ and $\Theta^{ij}_{mn} (\theta,\phi)=\left( 3\hat{r}^i\hat{r}_m-\delta^i_m\right)\left( 3\hat{r}^j\hat{r}_n-\delta^j_n\right)$.
Similarly, for $\corr_3$, one finds
\begin{equation}
\E[ b^i(\boldsymbol{r},t_1)  b^{j}(\boldsymbol{r},t_1) b^{k}(\boldsymbol{r},t_2) b^{l}(\boldsymbol{r},t_2)]= R_4(r) \Theta^{ijkl}_{mnpq} (\theta,\phi) \E[\hat{\mu}^m(t_1)\hat{\mu}^n(t_1)\hat{\mu}^p(t_2)\hat{\mu}^q(t_2)]
\end{equation}
For convenience, we define a new frame (XYZ) with the precession axis as the Z-axis so
\begin{align}
\hat{\mu}_X&=\hat{\mu}_\perp \cos(\int\mathrm{d}t\omega_L + \varphi)\\
\hat{\mu}_Y&=\hat{\mu}_\perp \sin(\int\mathrm{d}t\omega_L + \varphi)\\
\hat{\mu}_Z&=\hat{\mu}_\parallel
\end{align}
in which $\hat{\mu}_\perp$, $\hat{\mu}_\parallel$, and $\varphi$ define the initial state at $t=0$, which is assumed random, and $\omega_L=\omega_{L,0}+\delta\omega_L$. To illustrate how we get the damping cosine function in Eq.~\ref{equ:corr2} and Eq.~\ref{equ:corr3}, we show how we calculate $\E[\hat{\mu}_X(t_1)\hat{\mu}_X(t_2)]$:
\begin{align}
    \E[\hat{\mu}_X(t_1)\hat{\mu}_X(t_2)]&=\E[\hat{\mu}_\perp^2\cos(\int^{t_1}_0\mathrm{d}t\omega_L + \varphi)\cos(\int^{t_2}_0\mathrm{d}t\omega_L + \varphi)]\\
    &=\frac{1}{3}\E[\exp\left[i\left(\omega_{L,0} (t_2-t_1)+\int^{t_2}_{t_1} \mathrm{d}t\delta \omega_L(t) \right)\right] +H.C.] \label{equ:ini}\\
    &=\frac{2}{3} \cos\left(\omega_0 T\right) \exp\left[-\frac{1}{2}\sigma T\right],  \label{equ:noi}
\end{align}
obtained by first averaging over initial states (Eq.~\ref{equ:ini}) ($\int \hat{\mu}_\perp^2 \sin\vartheta \mathrm{d}\vartheta =\frac{4}{3} $, $\int \exp\left(2 i \varphi\right) \mathrm{d}\varphi =0$) and then the noise in $\omega_L$ (Eq.~\ref{equ:noi}), in which $\int^{t_2}_{t_1}\mathrm{d}t\delta \omega_L(t)$ has a Gaussian distribution of $N\left(0, \sqrt{\sigma T}\right)$. A similar calculation can be carried out for other terms and all terms in the new frame can be converted back to the molecule frame with a rotation matrix. All together, we can get $\E[\hat{\mu}^m(t_1)\hat{\mu}^n(t_2)]=G^{mn}\cos\left(\omega_0 T\right) \exp\left[-\frac{1}{2}\sigma T\right]$, where $G^{mn}$ is a matrix of constants reflecting the symmetry of our system. Plugging this back to Eq.~\ref{equ:corr2b}, we get $g^{i,j}(\theta,\phi) \equiv \Theta^{ij}_{mn} (\theta,\phi) G^{mn}$. Similarly, for $\corr_3$, $\E[\hat{\mu}^m(t_1)\hat{\mu}^n(t_1)\hat{\mu}^p(t_2)\hat{\mu}^q(t_2)]=2G^{mnpq} \E[\exp\left[i\left(2\omega_{L,0} (t_2-t_1)+2\int^{t_2}_{t_1} \mathrm{d}t\delta \omega_L(t) \right)\right] +H.C.]=G^{mnpq}\cos(2\omega_{L,0}T)\exp\left[-2\sigma T\right]$.

\section{Additional data/Simulations}
\begin{figure}[b]
    \centering
    \includegraphics[width=\linewidth]{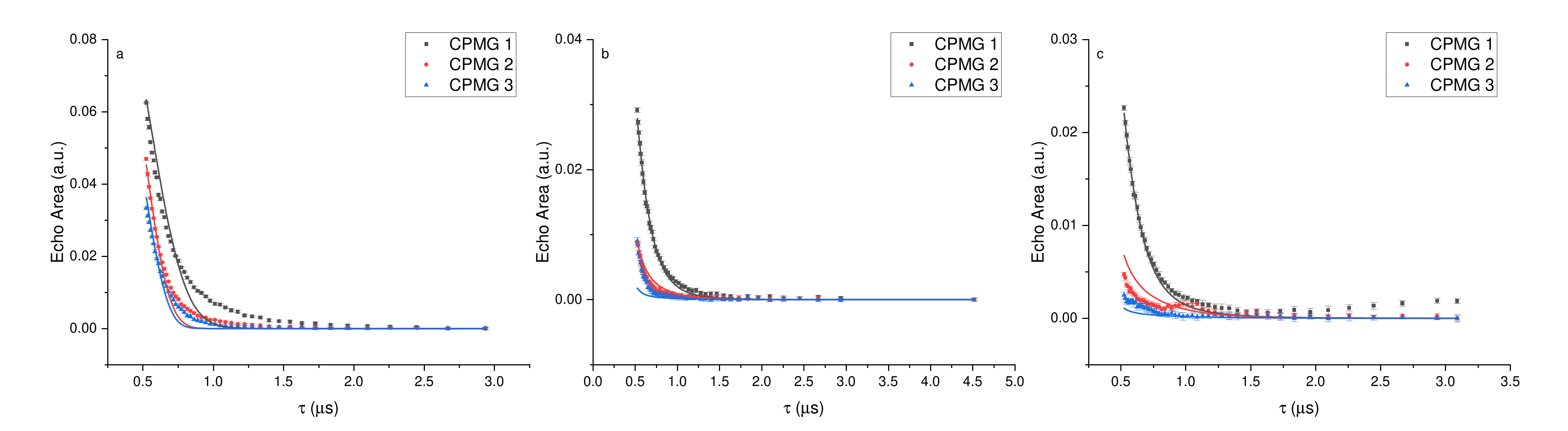}
    \caption{Additional CPMG data (points) and fits (solid lines) at fields of (a) 25, (b) 50, and (c) 75~Oe.}
    \label{fig:lower_field}
\end{figure}

\begin{figure}[h]
    \centering
    \includegraphics[width=\linewidth]{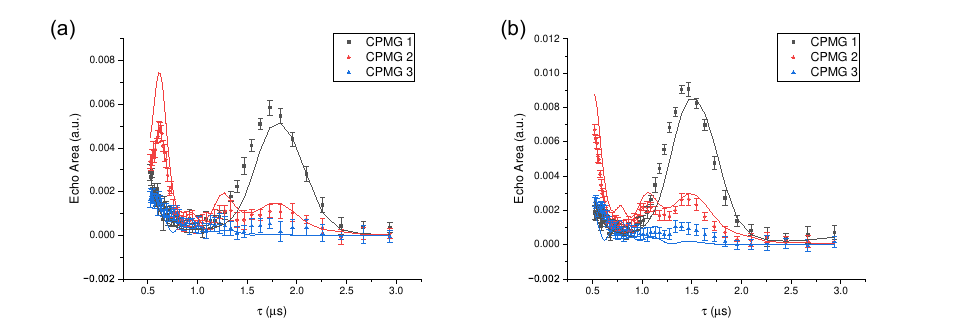}
    \caption{Additional CPMG data (points) and fits (solid lines) at fields of (a) 125 and (b) 150~Oe.}
    \label{fig:mid_field}
\end{figure}

\begin{figure}[h]
    \centering
    \includegraphics[width=\linewidth]{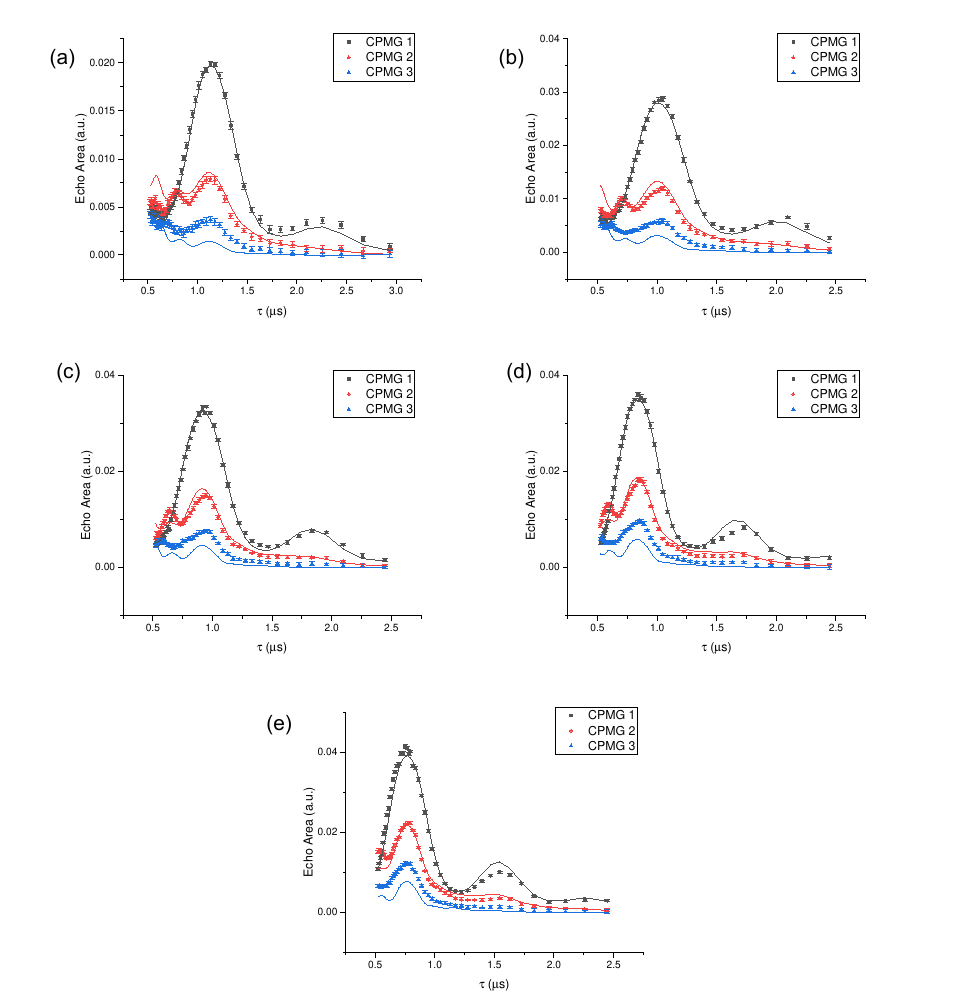}
    \caption{Additional CPMG data (points) and fits (solid lines) at fields of (a) 200, (b) 225, (c) 250, (d) 275, and (e) 300~Oe.}
    \label{fig:higher_field}
\end{figure}

\begin{figure}
    \centering
    \includegraphics[width=\linewidth]{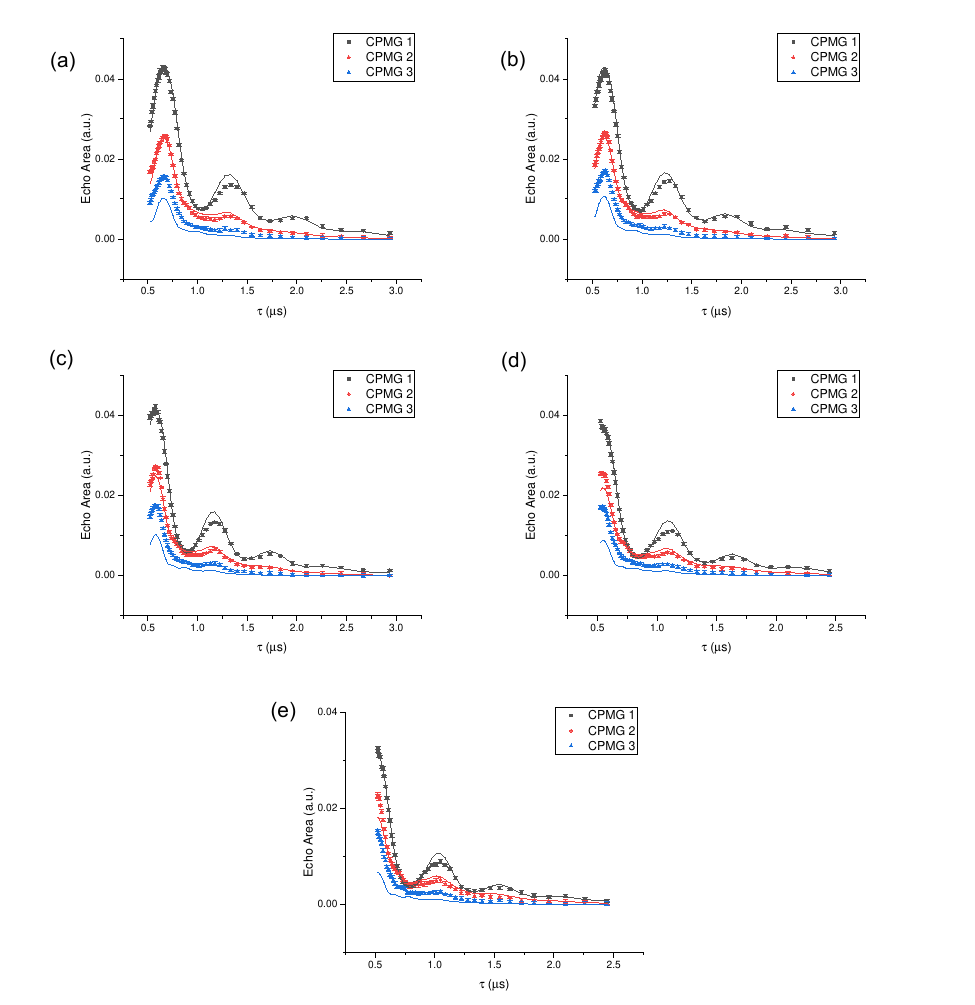}
    \caption{Additional CPMG data (points) and fits (solid lines) at fields of (a) 350, (b) 375, (c) 400, (d) 425, and (e) 450~Oe.}
    \label{fig:high_field}
\end{figure}

Fig.~\ref{fig:lower_field}--\ref{fig:high_field} show the data at other field not shown in the main text but used in the fitting.  %

\section{Fitting without noise in $E$}

\begin{figure}[t]
    \centering
    \includegraphics[width=.8\linewidth]{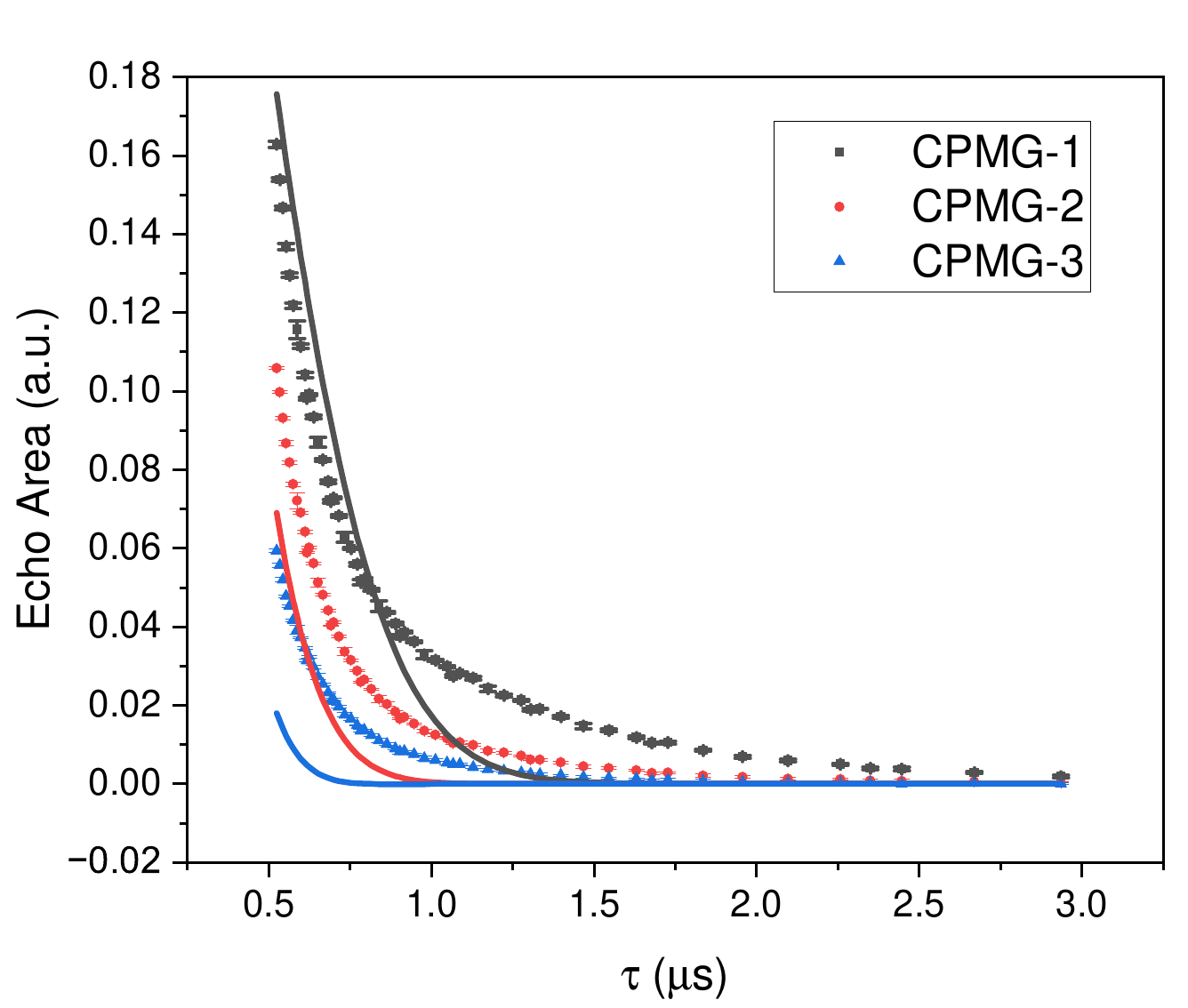}
    \caption{The result of fitting the zero-field data after forcing $S_E=0$.  The fit is substantially worse in comparison  to when $S_E$ is included in the model, cf.~Fig.~10(a) in the main text.}
    \label{fig:noE}
\end{figure}

As presented in the main text, our noise/decoherence model includes field noise due to a nuclear spin bath as well as noise in the anisotropy parameter $E$, $S_E\propto1/f$. This noise in $E$ presents a limiting factor on decoherence at the CT, where the effects of field fluctuations are largely filtered out.  At zero field, where the CT occurs, the only noise sources in our model that can lead to decoherence are $S_{\boldsymbol{B}_{I}\otimes\boldsymbol{B}_{I}}$ and $ S_E$ since $S_{B_I}$ is filtered completely by the CT.  This raises the question of whether the second-order field fluctuations $S_{\boldsymbol{B}_{I}\otimes\boldsymbol{B}_{I}}$ could be sufficient to account for our observed decoherence at zero field.  Re-running a fit of our model at zero field, forcing $S_E=0$ while allowing all other parameters to vary~\footnote{Excluding $S_{B_I}$ since the simulations have no dependence on it at zero field.}, produces the results shown in Fig.~\ref{fig:noE}.  The fit is far worse than what we obtain by including $S_E$, as shown in Fig.~10(a) in the main text.  This strongly suggests that the fluctuations in $E$ play an essential role in decoherence at the CT and that second-order fluctuations are insufficient to account for the observed zero-field behavior.

\section{discussion of coherent model on intermediate range of field}
\begin{figure}[t]
    \centering
    \includegraphics[width=0.95\linewidth]{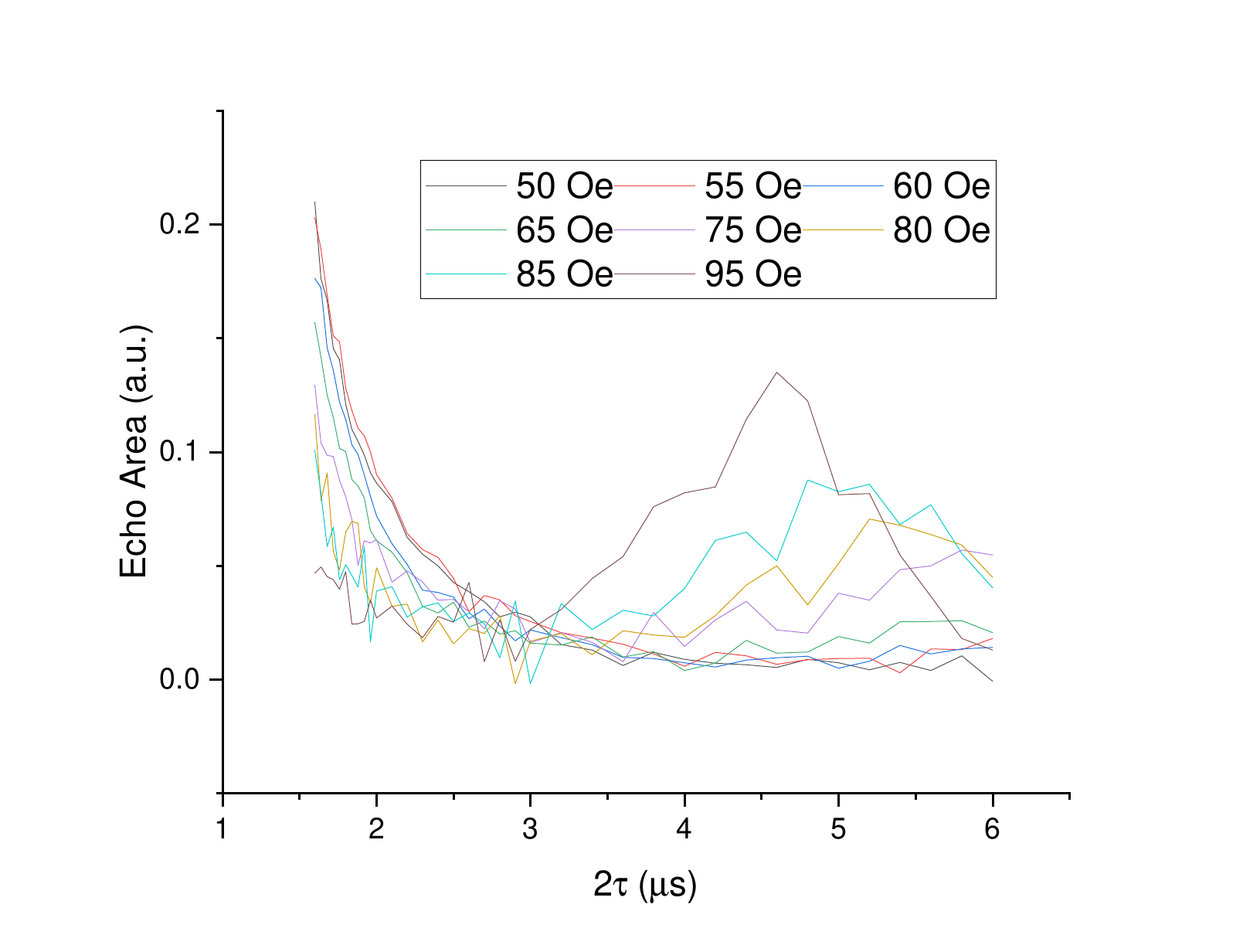}
    \caption{Hahn echo area at intermediate fields, 50 -- 95~Oe. The signal appears to revive after seemingly fully decohering. Data was taken with $5$\% dilution \textbf{2} toluene solution at $1.9$~K and $4639$~MHz.
    }
    \label{fig:Echo_CT}
\end{figure}

\begin{figure}[t]
    \centering
    \includegraphics[width=0.95\linewidth]{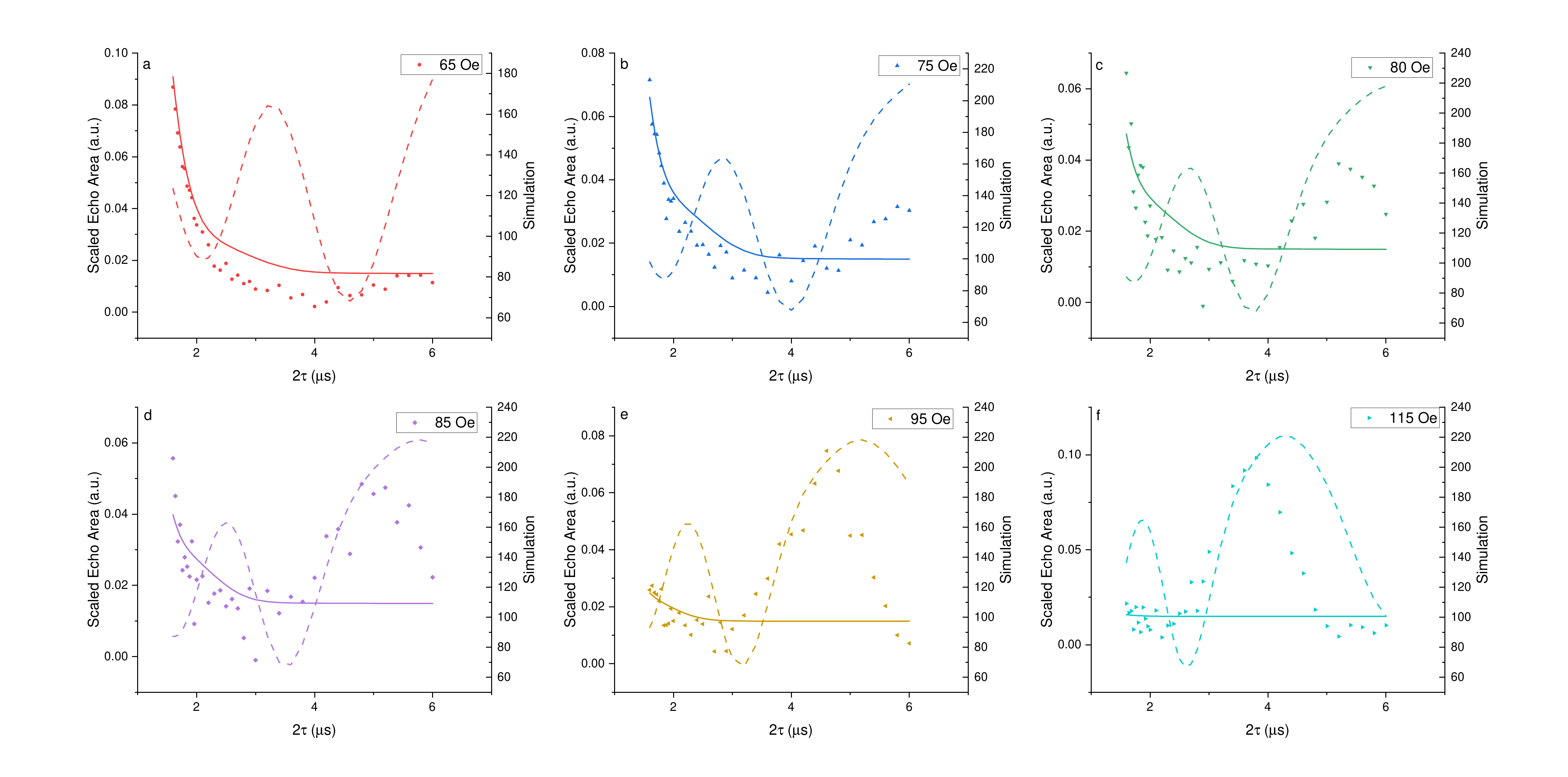}
    \includegraphics[width=0.95\linewidth]{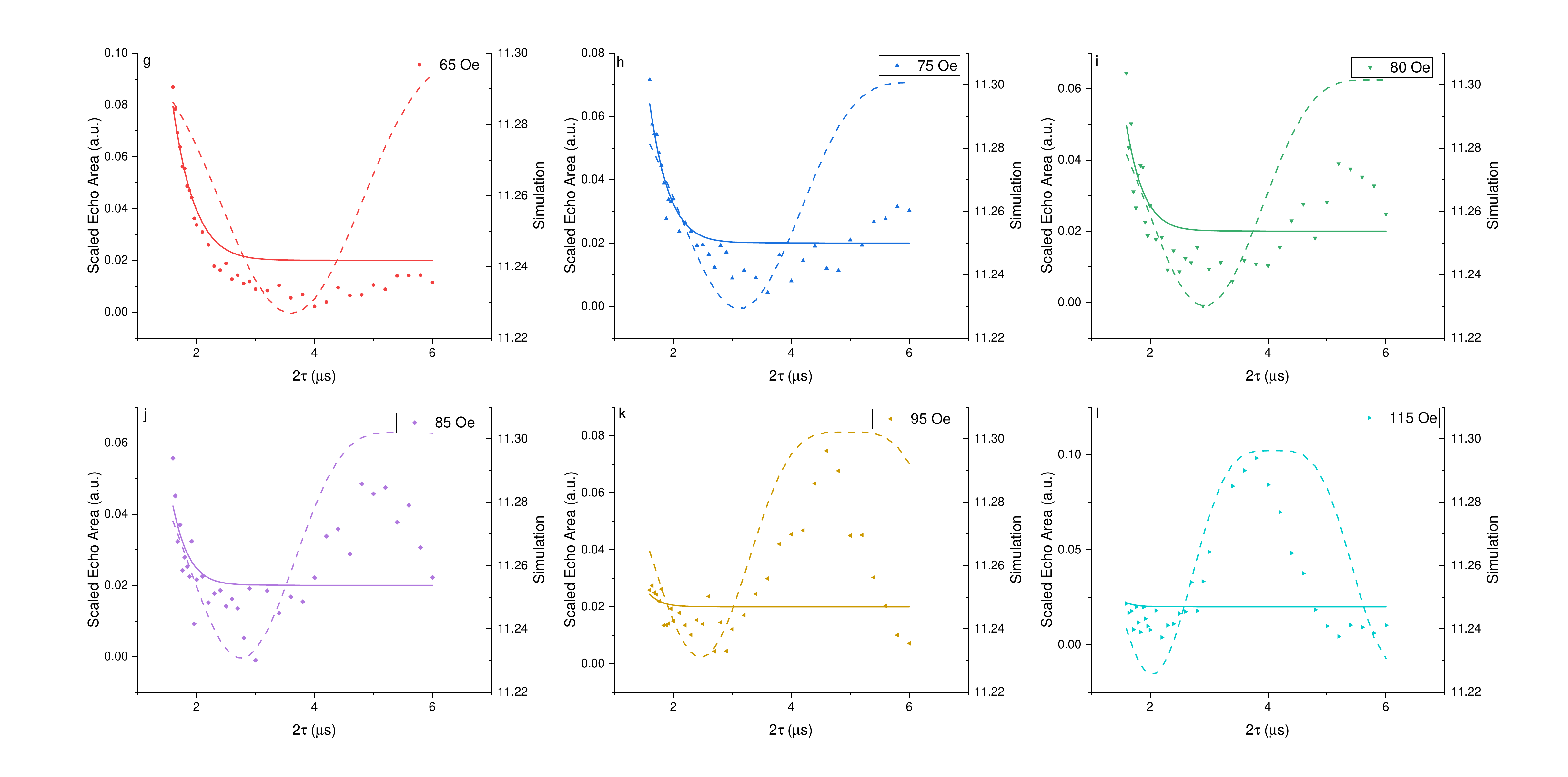}
    \caption{Attempted fitting at intermediate field values with hyperfine coupling. Panels a--f show the results from the same procedure used for Fig.~9, as described in main text, except allowing for refitting the global scaling parameter and offset. Note that with these hyperfine coupling parameters, the undecohered waveform (dashed line) shows additional structure with small $\tau$ not present in the data. Panels g--l are the results from fitting using a model of coupling to multiple ($n_p$) protons with small values of hyperfine coupling, Eq.~\ref{eqn:ham_eseem_n}. The modulation frequency is largely the nuclear Larmor frequency as a result.}
    \label{fig:fit_expl_medium}
\end{figure}

For intermediate-range fields, there is a revival of the signal after the system has seemingly fully decohered. Fig.~\ref{fig:Echo_CT} show the Hahn echo area in the range of $50$--$95$~Oe. This revival is in fact consistent with ESEEM oscillations. The echo area initially decays in a quasi-exponential way, but then recovers at later times, producing a signal that peaks at a time that is roughly consistent with the Larmor precession frequency.  We find that this behavior is not well produced with our coherent-model simulations, highlighting the advantage of modeling the environment as a noise source.  Figure \ref{fig:fit_expl_medium} illustrates the inability of the coherent-coupling model to adequately reproduce the data in this range of fields.  In the top six panels (a-f), we show experimental and simulated results for a sampling of six fields in this regime. The dashed lines represent the results of simulations without decoherence while the solid lines show the results of fitting.  For the fitting, we fixed the hyperfine parameters to those found from the fitting of data in the $195$ -- $395$~Oe field range and then refit the the data over all fields, allowing all other parameters ($T_2$ for each field, scaling parameter, offset) to vary.  There are two important failures of the simulations here:  1) Without decoherence, the simulations show a peak at values of $2\tau$ (e.g.~$\sim\!3\,\upmu$s for 75~Oe) where no peak is observed in the data.  This peak corresponds to oscillations at about twice the Larmor frequency, a known effect in ESEEM.  2) The fitting is unable to meaningfully reproduce the data.  One important observation is that the signal size in this range of fields is approximately an order of magnitude smaller than in the higher-field range.  In order to account for the small signal size as well as the rapid decay of low values of $\tau$ and the absence of the low-$\tau$ peak, the fit yields an extremely fast decay that does not account for the signal revival at longer values of $\tau$.  Attempts to force $T_2$ to be larger to allow for observation of the revival results in a signal amplitude that is vastly larger than the observations in this regime. 

To attempt to address the issues described above, we chose to reduce the hyperfine coupling strength by more than an order of magnitude: $A_{zz} = 0.2$~MHz and $A_{z\perp}=0$.  The motivation for doing this is to eliminate the low-$\tau$ peak in the simulations.  An additional consequence of reducing $A_{zz}$ to this small value makes the amplitude of the ESEEM oscillations much smaller than the observations.  To compensate for this, we allowed for coupling of the electronic spin to a large number of protons, not just one as in our previous model.  In other words, we update Eq.~2 (main text) to 
\begin{equation}
        \mathcal{H}=-D S_z^2+E (S_x^2-S_y^2)+g_s \mu_B\boldsymbol{B}\cdot\boldsymbol{S}
    +\sum_i\left( A_{zz} S_z I_{i,z}- g_p\mu_p\boldsymbol{B}\cdot\boldsymbol{I}_i  \right).
    \label{eqn:ham_eseem_n}
\end{equation}

For the small value of $A_{zz}$ we use, we find that the ESEEM oscillations retain their shape as we increase the number of protons $n_p$ in the sum and the oscillation amplitude scales linearly with $n_p$, which we treat as an additional fitting parameter.  Figure \ref{fig:fit_expl_medium}(g-l) show the results of an attempt to employ this model.  We indeed see that the additional low-$\tau$ peak is no longer visible in the damping-free simulations.  However, the fit still fails to adequately reproduce the experimental data, again showing a short $T_2$ to account for the initial fast decay of the signal.  Again, the fundamental issue here appears to be that longer values of $T_2$, while allowing the simulations to exhibit oscillations, produce signal sizes that are much too large to match the experimental data.  

\newpage
\bibliography{Cr7Mnbib.bib}